\documentclass[journal,12pt,onecolumn,draftclsnofoot,]{IEEEtran}

\usepackage{amsmath,amsfonts}
\usepackage{algorithm}
\usepackage{array}
\usepackage[caption=false,font=normalsize,labelfont=sf,textfont=sf]{subfig}
\usepackage{textcomp}
\usepackage{stfloats}
\usepackage{url}
\usepackage{verbatim}
\usepackage{graphicx}
\usepackage{cite}

\usepackage{amssymb}
\usepackage{subfiles}
\usepackage[nohyperlinks,nolist]{acronym}% no list para quitar la lista de abreviaciones
\usepackage{multirow}
\usepackage{tabularx}
\usepackage{booktabs}
\usepackage[inkscapeformat=png]{svg}
\usepackage{algpseudocode}
\usepackage{hyperref}
\usepackage{orcidlink}

\begin{document}

%\title{\textcolor{red}{Decentralized and Adaptive Task Offloading via On-Device \acl{DRL}: Trade-offs in Training and Energy Efficiency}}
\title{On-Device \acl{DRL} for Decentralized Task Offloading: Performance trade-offs in the training process}

\author{Gorka~Nieto\,\orcidlink{0000-0001-8258-5169},%
        Idoia~de~la~Iglesia\,\orcidlink{0000-0003-0799-9454},%
        Cristina~Perfecto\,\orcidlink{0000-0001-6729-6844},%
        and~Unai~Lopez-Novoa\,\orcidlink{0000-0002-2707-8946}%
\thanks{Gorka Nieto is with Ikerlan Technology Research Centre, Basque Research and Technology Alliance (BRTA), Arrasate, Spain, and also with School of Engineering in Bilbao, University of the Basque Country (EHU), Spain, (e-mail: gnieto004@ikasle.ehu.eus).}%
\thanks{Idoia de la Iglesia is with Ikerlan Technology Research Centre, Basque Research and Technology Alliance (BRTA), Arrasate, Spain (e-mail: idelaiglesia@ikerlan.es).}%
\thanks{Cristina Perfecto and Unai Lopez-Novoa are with School of Engineering in Bilbao, University of the Basque Country (EHU), Spain, (e-mails: cristina.perfecto@ehu.eus and unai.lopez@ehu.eus).}%
\thanks{This work has been partially supported by the COGNIT project (Grant agreement ID: 101092711), funded by the European Commision; and by the XWAVE research grant (KK2025/00056), funded by the Basque Government.
}%
}
        % <-this % stops a space
% \thanks{This paper was produced by the IEEE Publication Technology Group. They are in Piscataway, NJ.}% <-this % stops a space
% \thanks{Manuscript received April 19, 2021; revised August 16, 2021.}}

% The paper headers
% \markboth{Journal of \LaTeX\ Class Files,~Vol.~14, No.~8, August~2021}%
% {Shell \MakeLowercase{\textit{et al.}}: A Sample Article Using IEEEtran.cls for IEEE Journals}

% \IEEEpubid{0000--0000/00\$00.00~\copyright~2021 IEEE}
% Remember, if you use this you must call \IEEEpubidadjcol in the second
% column for its text to clear the IEEEpubid mark.

\maketitle

\begin{abstract}
%\textcolor{red}{[75-200 palabras]} 
Allowing less capable devices to offload computational tasks to more powerful devices or servers enables the development of new applications that may not run correctly on the device itself. Deciding where and why to run each of those applications is a complex task. Therefore, different approaches have been adopted to make offloading decisions. In this work, we propose a decentralized \ac{DRL} agent to address the selection of computing locations. Unlike most existing work, we analyze it in a real testbed composed of various edge devices running the agent to determine where to execute each task. These devices are connected to an \ac{MEC} server and a Cloud server through 5G communications. We evaluate not only the agent's performance in meeting task requirements but also the implications of running this type of agent locally, assessing the trade-offs of training locally versus remotely in terms of latency and energy consumption.
\end{abstract}

\begin{IEEEkeywords}
\acf{IoT}, Edge-Cloud-Continuum, Task Offloading, \acf{DRL}, 5G.
\end{IEEEkeywords}

\section{Introduction}
\label{sec:intro}

As the \ac{IoT} expands into more domains, such as smart cities, healthcare, or industrial automation, the number of interconnected \ac{IoT} devices that sense, process data, and communicate with each other continues to grow rapidly. This leads to an ever-increasing volume and heterogeneity of data being generated, processed, and exchanged \cite{Al-Ali2024}. Moreover, these devices often face a critical issue: energy consumption. Most \acs{IoT} nodes are resource-constrained in terms of limited processing capacity. Still, they usually rely on limited battery power or energy harvesting, which directly impacts their operational lifetime and scalability \cite{Cook2023}.

Due to the limitations above, \ac{IoT} devices may sometimes be unsuitable, and other alternative device types may be required. For instance, to process the vast amount of data generated by \ac{IoT} devices and create more valuable services, integrating \ac{IoT} with \ac{CC} has been opening up new solutions in recent times \cite{TyagiHimaniandKumar2020}. However, for specific situations where latencies are critical, the \ac{CC} paradigm is not suitable. Therefore, \ac{MEC} paradigm came to overcome the limitations of both \ac{IoT} and \ac{CC} concepts, by expanding the capacity of devices in terms of computing resources, and reducing latencies due to distance, respectively, which enables more applications in other use cases, such as healthcare, smart cities or latency-sensitive multimedia communication, among others \cite{Khattak2025}. 

The combination of these paradigms creates an \ac{IoT}-Edge-Cloud continuum. This unified system integrates \ac{IoT} devices, edge, and cloud resources, enabling the distribution of computational tasks across heterogeneous devices, from resource-constrained \ac{IoT} nodes to powerful cloud servers. As a drawback, it also introduces significant challenges in managing heterogeneity, distributed data, or scalability \cite{Gkonis2023}.

Thus, task offloading has emerged as a highly effective strategy to optimize resources, minimize delays, and consumption, among other goals. Furthermore, as \acs{IoT} devices are required to run increasingly demanding applications, there is a growing need for intelligent task offloading mechanisms in edge devices \cite{Xiao2022}. Moreover, as conditions can change due to the dynamic nature of common industrial environments (such as wireless communications and applications with varying requirements), there is a need for tools that can adapt to these changes. That is precisely why \ac{ML} techniques, and more specifically, \ac{DRL} techniques, have become so relevant in this topic, thanks to their adaptability to changing conditions \cite{Zabihi2023}, resulting in \ac{AI} agents that can act autonomously.

Furthermore, decentralized approaches enable autonomous decisions, in contrast to centralized approaches, resulting in higher availability and fault tolerance, while avoiding the risk of single-point failures and connectivity issues \cite{Zheng2025}. On-device \ac{ML} refers to running ML tasks (inference or training) directly on a device, without relying on external servers or cloud services. While frameworks like TensorFlow Lite are optimized for inference in embedded devices with an operating system, there is a need for tools that can streamline the entire development process, from model training and optimization to deployment and monitoring, as these tools are not optimized for training \cite{Anwar2024}. 

%Hence, %unlike many existing works that rely on inference-only frameworks such as \textit{TensorflowLite} or \ac{ONNX}, 
%in this work, we %employ TensorFlow to enable on-device training and real-time policy adaptation by implementing 
Hence, in this work, we analyze the behavior and performance of on-device \ac{DRL} agents for computation offloading, with emphasis on the training process. To that end, we implement the whole process of a \ac{DRL} model on-device, enabling the \ac{AI} agent to learn online. This approach enables a precise evaluation of the practical trade-offs associated with on-device \ac{DRL} agent training, demonstrating its feasibility as well as its benefits and limitations in comparison to centrally trained, distributed inference schemes operating under dynamic, real-world conditions. Our results show that, using a lightweight \ac{DRL} model, on-device training introduces reasonable computation time and energy consumption, obtaining better overall performance than offloading the training process to a remote location.

The remainder of the paper is organized as follows. Section \ref{sec:related_work} provides a summary of the related work. The problem formulation and the proposed solution are presented in Section \ref{sec:system_model}, which are evaluated in the context of the real-world setup described in Section \ref{sec:setup}. Section \ref{sec:results} presents the analysis of the results, and finally, Section \ref{sec:conclusion} provides insights into this work, along with potential directions for future research.

\section{Related Work}
\label{sec:related_work}

Task offloading has been widely studied using diverse decision techniques, including traditional optimization methods \cite{Wang2014}, heuristic \cite{Mei2023} and metaheuristic \cite{Lakhan2024} approaches, control-theoretic approaches \cite{Ling2024}, game-theoretic approaches \cite{Xu2022}, \ac{ML}-based (e.g. \ac{SL} \cite{Yu2017} or \ac{RL} \cite{Moghaddasi2024}) approaches, and hybrid approaches that combine multiple paradigms \cite{Choppara2024}. While traditional optimization methods ensure accuracy, they are unsuitable for large-scale and dynamic environments due to their computational complexity \cite{Bian2024}. Heuristic and metaheuristic approaches provide faster solutions but are typically static and problem-specific, with limited ability to adapt to changing network states \cite{Wang2021}. Control-theoretic methods are effective in continuous dynamics but require accurate system models that are unfeasible to model due to real-world uncertainties\cite{Pedrosa2021}. In contrast, game-theoretic approaches often rely on strong assumptions about agents' knowledge and are computationally demanding \cite{Vainer2021}. In contrast, \ac{DRL} offers a model-free and adaptive framework in which each \ac{AI} agent learns a policy to maximize a reward defined around the optimization objective \cite{dong2020deep}.

While many approaches, such as \cite{Shao2025}, propose centralized decision-making \ac{DRL} approaches, distributed or decentralized \ac{DRL} approaches are becoming increasingly relevant, as seen in \cite{Hevesli2025}. In these works, despite their decentralized execution during inference or data collection, most of these approaches still rely on centralized training paradigms. In such setups, experiences gathered from distributed \ac{AI} agents are aggregated and used to update a shared policy or value function, typically maintained on a central server or coordinator. Additionally, centralized training necessitates the frequent transmission of experience to a central server, which can introduce latency and become a bottleneck—especially in scenarios involving many agents \cite{Provatas2025}. Additionally, wireless networks commonly used in edge setups are often unstable, with issues like packet loss and fluctuating bandwidth that hinder synchronization and data consistency. 

Despite the growing interest in decentralized/distributed computation offloading and edge-based \ac{DRL}, very few studies address the practical challenges of running such systems on real edge devices, such as the \textit{Jetson Nano} or \textit{Raspberry Pi}. Even fewer works provide empirical evaluations in realistic setups, leaving a gap in understanding how these systems perform under actual constraints, such as limited resources and unstable connectivity. One example of this is the work presented in \cite{Qiu2021}, which provides a practical application of \ac{ML} techniques for offloading decisions. It implements a distributed and collective \ac{DRL} algorithm that leverages knowledge from various \ac{MEC} environments. The implementation relies on a stable Wi-Fi connection within a static indoor setting, thereby minimizing the impact of channel fluctuations on the model and eliminating the need for further analysis in dynamic environments.

Regarding on-device \ac{ML}, several studies deploy pre-trained models using lightweight inference engines such as \textit{\ac{ONNX}} or \textit{TensorFlow Lite}, which are well-suited for constrained devices but do not support on-device training, which limits their adaptability \cite{Zhu2024}. This limitation, along with other on-device training challenges, brings the need for more flexible and powerful \ac{ML} frameworks that can support both inference and training on edge devices. For instance, the work in \cite{Sayeedi2024} investigates the use of high-level (\textit{PyTorch}) and low-level (\textit{LibTorch}) libraries for \ac{ML}-based image classification on \ac{IoT} devices. They evaluate the models' performance and quantify the green footprint of \ac{ML} model training, including metrics such as execution time, memory usage, power consumption, and CPU temperature.

In light of this, despite the interest in deploying \ac{DRL} at the edge to maintain the adaptability of these models, particularly in the context of computation offloading, real-world evaluations of locally trained agents remain scarce. Most existing studies focus on theoretical models or simulations, overlooking the practical constraints and trade-offs encountered during online training in unstable, real-world deployments. In such settings, models can misbehave due to the environment's unpredictability, which may be ignored in controlled simulated environments. Furthermore, the comparative analysis between centralized training and on-device learning under these conditions is missing. To address this gap, this work implements intelligent offloading strategies that explicitly account for the costs and benefits of training locally, especially in real environments where connectivity is intermittent or computational resources are limited. This approach enables continuous, on-device learning, supporting real-time decision-making where centralized training is impractical or suboptimal.

% \cite{Li2025}

% \cite{Chandrinos2024}

% \cite{Ivanov2025}

% \cite{Wang2024}

% \cite{AlAidaros2023}

% \cite{Artemenko2021}

% \cite{Zhao2023}

% \cite{BenSada2024}

% \subsection{\ac{ML} compression techniques}
% \cite{Yahmed2023}: DRL systems' high computational and energy costs prevent their direct deployment on platforms with low processing power. Generally non-trivial and more difficult compared to vanilla DL deployment. Quantization [14, 151, for instance, is more challenging in DRL and may hinder the policy's long-term decision-making since the agent's current action strongly affects its future states and actions [16
% TF Lite [30] and Core ML [31] employ model compression strategies such as quantization before deploying DRL models to edge platforms.
% Challenges: Agent Ezport. converting its trained model into the required format or by (2) using dedicated frameworks to transform the model to a format that runs on the deployment platform. 
% Model Conversion, It includes issues associated with any setting misbehavior or incorrect use of model conversion frameworks (ONNX)). Model Quantization. Quantization lowers the accuracy of model parameters' representation in order to minimize the memory footprint of DNNs. In this challenge, developers often struggle with combining quantization frameworks like TF Lite with Other frameworks or platforms or working with varied precision floating points  DRL specific deployment challenges remain unstudied

\section{Problem Formulation and Evaluated Algorithms}
\label{sec:system_model}

\begin{figure}[H]
    \centering
    \includegraphics[width=0.5\linewidth]{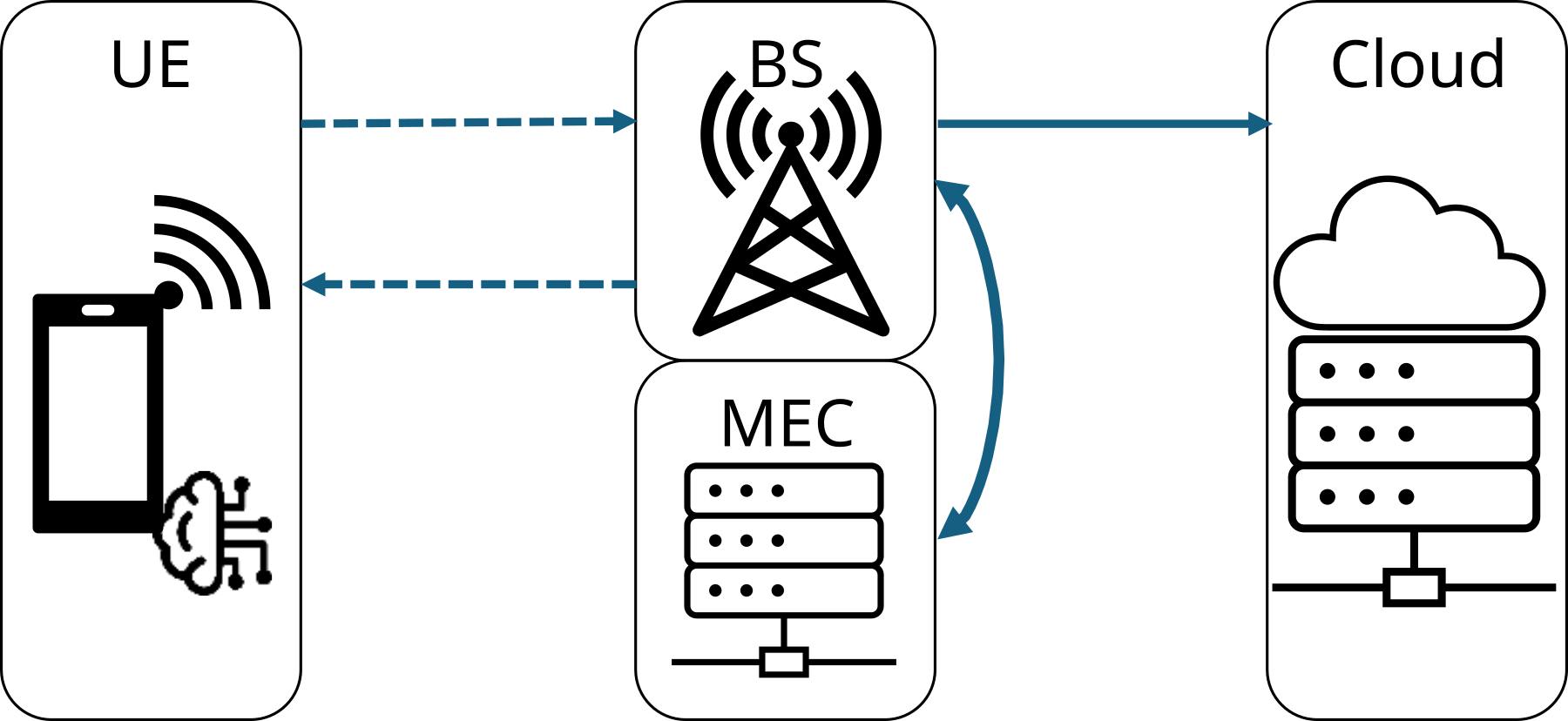}
    \caption{Entities involved in the system model}
    \label{fig:system}
\end{figure}

In this section, we present the optimization problem to be solved and the tools we use to address it. In this work, different \acp{UE} must decide whether to run incoming computing tasks locally or offload them to the \ac{MEC} or cloud server (see Figure \ref{fig:system}). Thus, apart from computing these incoming tasks, they must also run an instance of an offloading decision algorithm, which they use to choose whether to run or offload the tasks. 

We begin by describing the available key metrics that the \ac{UE} can gather, which the algorithm analyzes to make informed decisions. Next, we outline the actions the algorithm can take and then explain the optimization goal. Following this, we introduce the evaluated \ac{DRL} algorithms and the training alternatives in this approach.

\subsection{Problem formulation}
\label{sec:prob_form}

A \ac{DRL} agent is implemented on each \ac{UE}. The \ac{UE} continually monitors specific parameters in real-time, which serve as inputs for the \ac{DRL} algorithm's decision-making process. 

The problem is modeled as a \ac{MDP} model. At every timestep $i$, a new task $(\Psi)$ arrives, and the \ac{DRL} agent evaluates the current state $s_i$. The state is defined by a  5-tuple vector space, as shown in Equation \ref{state}:
\begin{equation}\label{state}
    s_i=\lbrace R_{i}^{n},R_{i}^{m},T_{i}^{n},C_i,\delta_i\rbrace
\end{equation}

\noindent where $R_{i}^{n}$ and $R_{i}^{m}$ represent the current available resources in the \ac{UE} and the \ac{MEC} server, respectively; $T_{i}^{n}$ represents the throughput experienced by the \ac{UE} from the attached \ac{BS}; and $C_i$ and $\delta_i$ represent the computing size and latency constraints of the incoming task, respectively. 

As retrieving metrics from a server in the Cloud could imply high latencies, we consider it to have sufficient compute resources in any scenario, and also to have multiple backup servers to serve the client at any time. Therefore, this information is not taken into consideration when making a decision. Hence, the selected state variables represent a minimal yet sufficient set of real-time metrics that can be reliably monitored on resource-constrained devices. Historical or long-term statistics are excluded to avoid additional memory overhead and to preserve the feasibility of continuous on-device learning. 
 
Considering the previous parameters, the \ac{DRL} agent must decide where to deploy and run each incoming task. This action represents how an incoming task shall be executed and is expressed as $\alpha_i \in \lbrace0,1,2 \rbrace$, being $\alpha_i=0$  local execution, $\alpha_i=1$ offloading to the \ac{MEC} server, and $\alpha_i=2$ offloading to the cloud server.

We define the performance metric as the energy consumption of the \ac{UE} conditioned by the success of the computing task. The success or failure of a computing task is evaluated by whether it fulfills its latency requirement. If this condition is not met, the performance will incur a penalty value $\eta$. It is essential to note that total latency encompasses decision time, transmission and reception time (if the data is transmitted), and processing time, all of which are affected by the parameters comprising $s_i$ and the chosen action $\alpha_i$.

The reward function for the \ac{DRL} agent $r$ is thus equal to the performance value, which is shown in Equation (\ref{qoe_calculation_only_energy}).

\begin{equation}\label{qoe_calculation_only_energy}
            r_i= \left\{ \begin{array}{ll}
        -e_{i} & \mbox{, if success} \\
        \eta & \mbox{, if fail} \\
        \end{array}
        \right.    
\end{equation}

This reward formulation strikes a balance between energy efficiency and task reliability, discouraging offloading decisions that may violate latency constraints under dynamic network conditions. As such, it reflects practical deployment requirements in real-world edge environments. 

The objective of the optimization problem is to maximize the overall reward (Equation \ref{qoe_calculation_only_energy}), defining the problem stated in Equation (\ref{optimization_problem}):

\begin{equation}\label{optimization_problem}
    \begin{array}{rrclcl}
        \displaystyle \max_{\alpha_i} & \multicolumn{3}{l}{\displaystyle \sum_{i=0}^{T-1}r_i} \\
        \textrm{s.t.} & & & \\
        c_1: & \alpha_i  & \in & \lbrace 0,1,2 \rbrace &, \forall i \in I\\
    \end{array}
\end{equation}

\noindent where $c_1$ is an action constraint indicating that each task can only be executed by the \ac{UE} itself, offloaded to the \ac{MEC} server, or offloaded to the cloud server.%; \textcolor{red}{$C2 - C4$ resource constraints mean that the dedicated computation resources to execute task $i$ cannot be higher than the total computation resources of the \ac{UE}, the \ac{MEC} and the cloud server respectively. }

In summary, Equation \ref{state} represents the parameters considered by the algorithm when deciding on an action to optimize the performance as defined in Equation \ref{qoe_calculation_only_energy}. In essence, the 5-tuple vector $s_i$ provides the inputs that enable the algorithm to make informed decisions, directly impacting the performance outcomes.

\subsection{DRL Algorithms}

Due to their complementary trade-offs between stability and computational complexity, two different \ac{DRL} algorithms are implemented and explained in this subsection: \ac{AC}, which combines both policy and value functions to enable more stable and efficient learning; and \ac{DQN}, which leverages value-based learning with \ac{DNN} to estimate action values. These algorithms have been validated in our previous work \cite{Nieto2024}.
    
\subsubsection{\acl{AC}}
\label{subsubsec:AC}
\begin{figure}[htbp]
  \centering
  \includegraphics[width=0.65\linewidth]{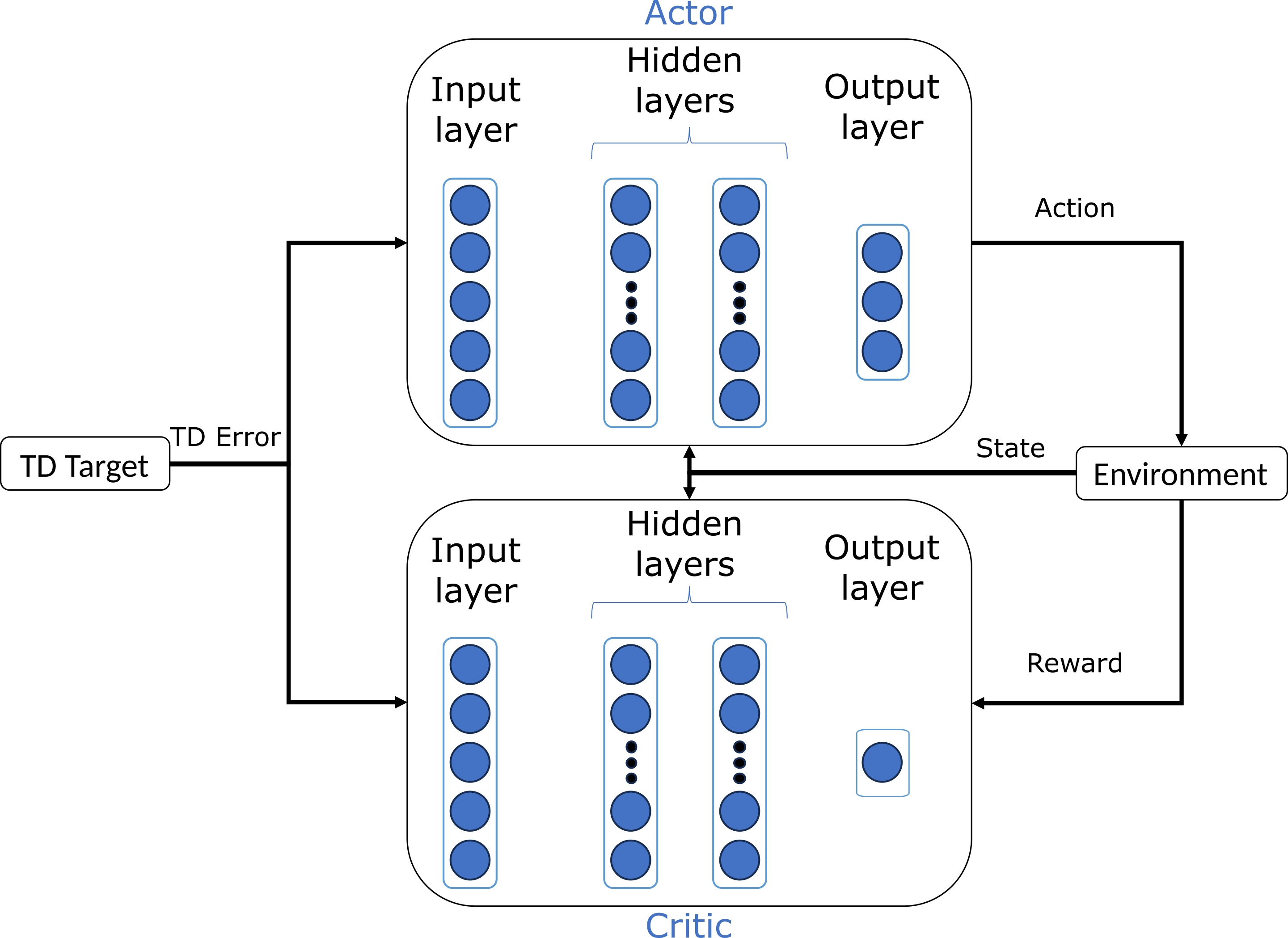}
  \caption{Architecture of the implemented \acl{AC} algorithm}
  \label{fig:algorithm}
\end{figure}

The \ac{AC} algorithm strikes a balance between policy-based and value-based methods, as illustrated in Figure \ref{fig:algorithm}.

This algorithm comprises $2$ networks (Actor and Critic) with a similar structure. Both the Actor and Critic networks receive the same $5$-dimensional state vector as input. Each of the Actor and Critic networks consists of:
\begin{itemize}
    \item \textbf{Input Layer}: accepts the 5-dimensional state vector.
    \item \textbf{1st Hidden Layer}: a fully connected (Dense) layer with $256$ neurons and \ac{TanH} activation.
    \item \textbf{2nd Hidden Layer}: another fully connected layer with $256$ neurons and \ac{TanH} activation.
    \item \textbf{Output Layer}: for the Actor, this layer is a fully connected layer with $3$ neurons (one for each possible action), followed by a softmax activation function to produce a probability distribution over possible actions. For the Critic, this layer is a fully connected layer with $1$ neuron, so the output is the scalar value estimate for the current state, with no activation function.
\end{itemize} 

The training is performed through online learning, where the agent interacts with the environment and updates its networks after each step, based on the observed reward and state transition, thereby eliminating the need for batches during training. The Critic is trained to minimize the \ac{TD} error, while the Actor is updated using the policy gradient method, weighted by the \ac{TD} advantage. Both networks are optimized using the Adam optimizer with a fixed learning rate of $ 1 \cdot 10^ {-5}$.

\subsubsection{\acl{DQN}}
\label{subsubsec:DQN}

\begin{figure}[htbp]
    \centering
    \includegraphics[width=0.65\linewidth]{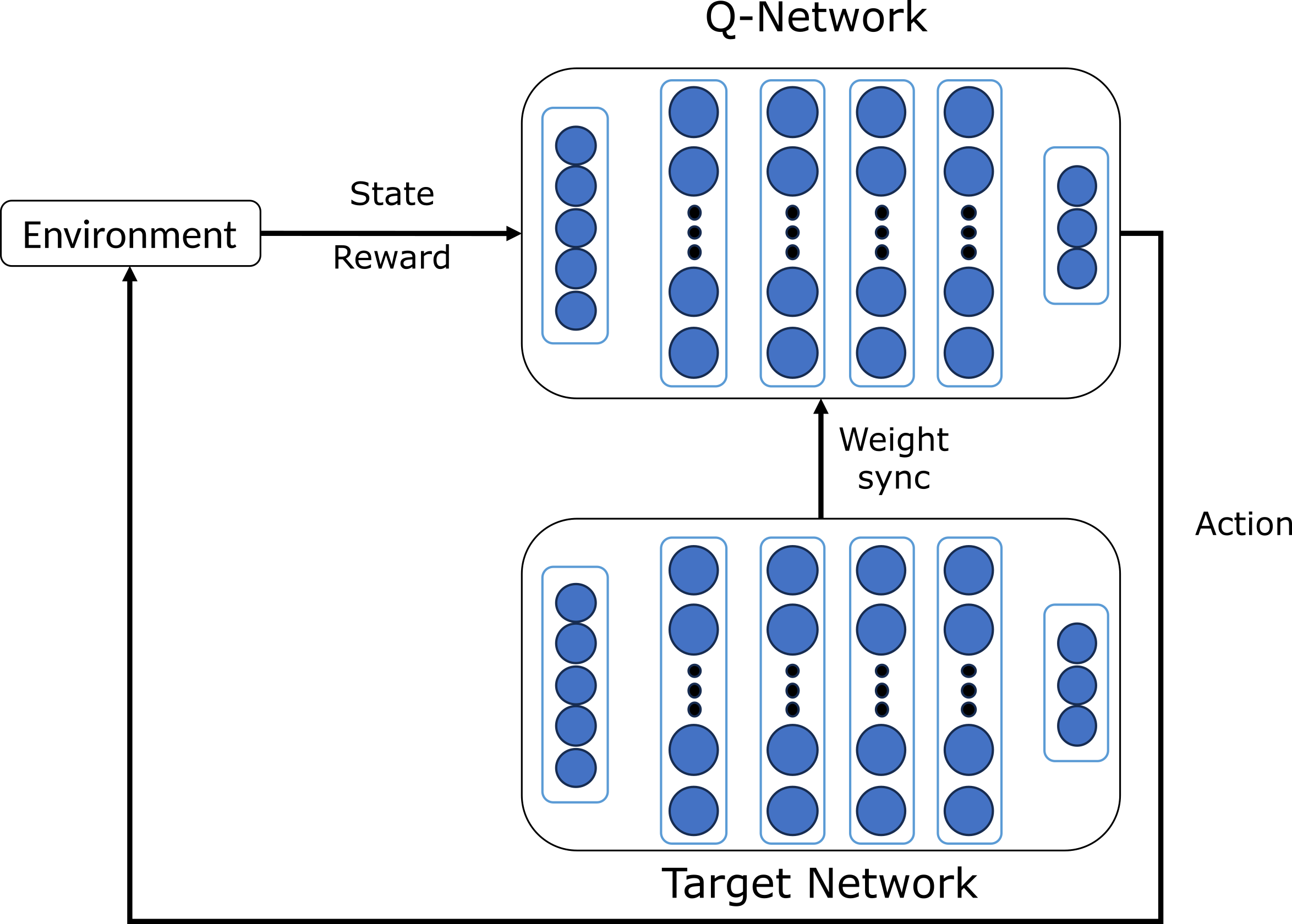}
    \caption{DQN model}
    \label{fig:DQN}
\end{figure}

The \ac{DQN} algorithm is a value-based \ac{RL} method that extends traditional Q-learning by utilizing a \ac{DNN} to approximate the Q-value function, thereby enabling it to handle larger state spaces. Its configuration is shown in Figure \ref{fig:DQN}.

The \ac{DNN} used to approximate the Q-value function takes the current state as input, and outputs Q-values for every possible action in the state. In this case, the neural network comprises four hidden layers, each containing 256 neurons, utilizing the \ac{ReLU} activation function. The produced are used to determine the action with the highest expected reward in the current state, following an epsilon-greedy policy. This policy balances exploration and exploitation by selecting a random action with a probability of $\epsilon$, while choosing the action with the highest Q-value otherwise. This model employs the Adam optimizer with a learning rate of $1 \cdot 10^{-5}$. 

Both algorithms are trained using an online, continual learning approach without large replay buffers or batch updates. This design choice reflects realistic edge deployments where memory availability is limited and timely adaptation is required.

\section{Experimental Setup}
\label{sec:setup}

In this section, we present the testbed used to conduct the empirical experiments described in the following section. First, we provide a detailed description of the testbed, followed by an explanation of the evaluation methods to analyze the performance of the on-device training process.

\subsection{Testbed}

\begin{figure}[hbt!]
    \centering
    \includegraphics[width=0.75\columnwidth]{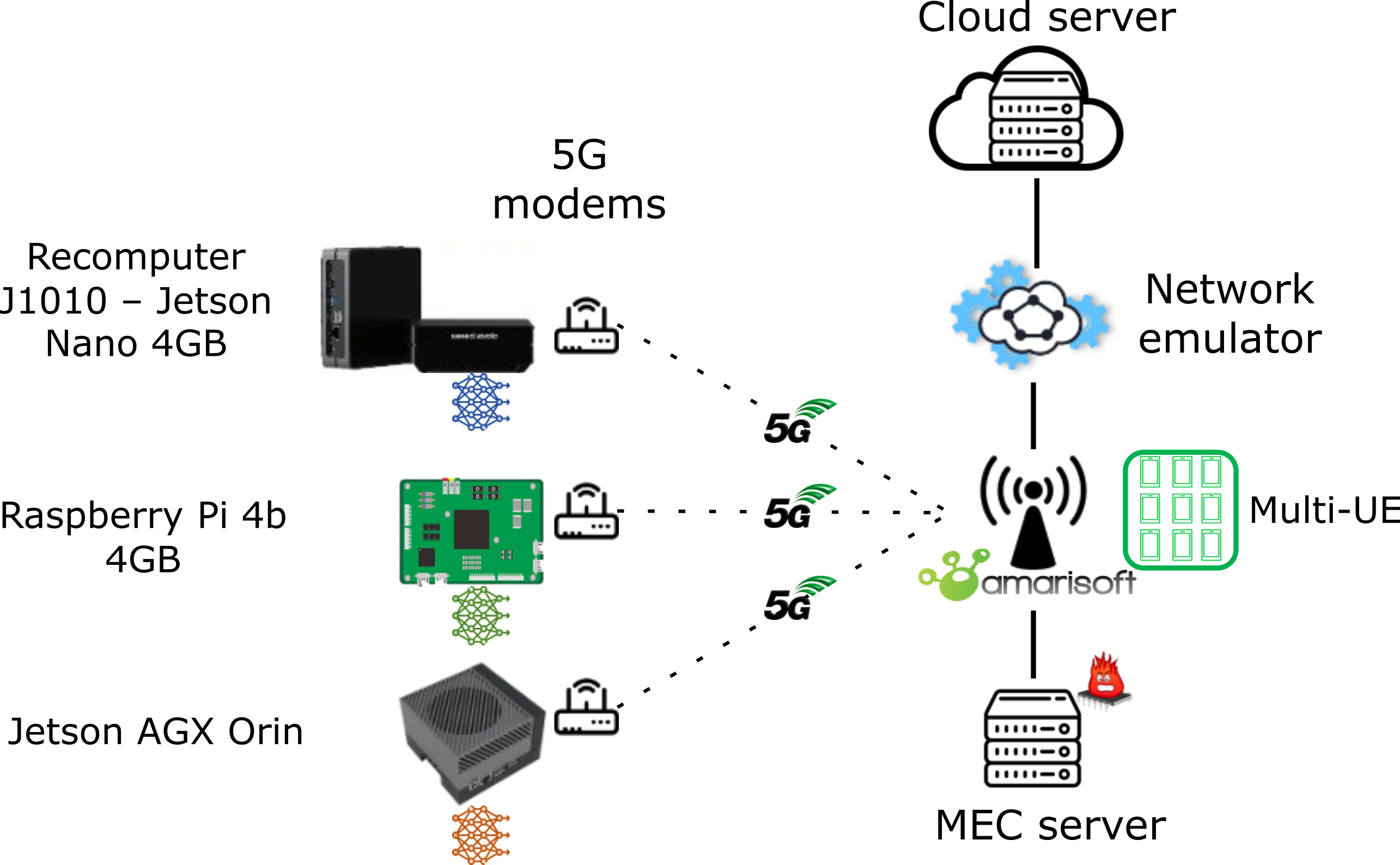}
    \caption{Architecture of the evaluation testbed}
    \label{fig:architecture}
\end{figure}

We created a testbed with the structure depicted in Figure \ref{fig:architecture} to conduct the empirical experiments explained in the following sections. %This testbed is a continuation of the one presented in \cite{Nieto2025}. 
As can be seen, we connect different \acp{UE} to a private 5G network using a 5G-capable modem for each of them, as they do not have integrated 5G communication capabilities. The 5G network is provided by a private \ac{BS}. The \ac{BS} is then connected to a \ac{MEC} server and to a Cloud Server. To emulate the network behavior required to reach the latter, we utilize an in-house developed network emulator. Finally, other emulated \acp{UE} are also connected to the same \ac{BS} to generate traffic that can affect the network status.

In this work, the \acp{UE}, which, as told in Section \ref{sec:prob_form}, must run an instance of an offloading decision agent, must also run a tool to obtain their energy consumption. We have implemented tools that provide real measurements and estimations, depending on the available tool for each \ac{UE}, which are described as follows: 

\begin{itemize}
    \item \textbf{Raspberry Pi Model 4}: equipped with a 1.5 GHz 64-bit quad core \textit{ARM Cortex-A72} processor and 4GB RAM, running \textit{Raspberry Pi} OS version 11. In this case, \textit{Powerjoular} \cite{Noureddine2022} is installed for energy consumption measurement, which provides reliable estimations of the CPU's energy consumption.
    \item \textbf{Recomputer J1010 / Jetson Nano}: powered by a 1.43 GHz quad-core \textit{ARM Cortex-A57} CPU and a 128-core\textit{ NVIDIA Maxwell} GPU, with 4GB LPDDR4 RAM and 16GB eMMC storage, running \textit{JetPack} 4.6.2 on the \textit{Jetson Nano} production module. The energy consumption is obtained using data read from the IIO Power Sensors\footnote{The Jetson Nano features an on-board INA3221 power monitor, exposed via the Linux IIO subsystem (sysfs). %See \href{https://docs.nvidia.com/jetson/archives/l4t-archived/l4t-3275/index.html#page/Tegra%20Linux%20Driver%20Package%20Development%20Guide/power_management_nano.html}{NVIDIA Jetson Power Management Guide}
    }
    \item \textbf{Jetson AGX Orin} features a 2.2 GHz 12-core \textit{ARM Cortex-A78AE} CPU and a 2048-core \textit{NVIDIA Ampere GPU} with 64 tensor cores, paired with 64GB LPDDR5 RAM and 64GB eMMC storage, running \textit{JetPack} SDK 6.5. The energy consumption is obtained by parsing the output of the \textit{tegrastats} utility, thereby obtaining the consumption from both the GPU and the CPU.
\end{itemize}

In order to provide 5G connectivity to the \acp{UE}, we use two different 5G modules:
\begin{itemize}
    \item \textbf{Quectel RM520N-GL}. A 5G NR module for global Sub-6 GHz SA deployments, offering peak downlink speeds up to $2.4$ Gbps and uplink up to $900$ Mbps in \ac{SA} mode. It supports 3GPP Release 16 and is ideal for high-throughput, low-latency applications. It is paired with the official EVB kit to enable USB connection with the \textit{Raspberry Pi} and with an M.2 adapter to the \textit{Jetson AGX Orin}.

    \item \textbf{SIMcom SIM8262E-M2}. A compact 5G module based on the \textit{Snapdragon X62} platform, optimized for \ac{SA} networks with support for peak downlink speeds of $2.4$ Gbps and uplink up to $1$ Gbps. It adheres to 3GPP Release 16 and is tailored for embedded applications requiring reliable 5G \ac{SA} connectivity. Its M.2 form factor supports USB 3.1, so it is used for connection with the \textit{Recomputer J1010}.

\end{itemize}

The 5G \ac{BS} is an \textit{Amarisoft} \ac{NR} \ac{SA} \ac{BS} (version 2024-09-04), a fully 3GPP Release 17-compliant solution, enabling detailed testing and deployment of Release 17 features. 

The servers run a Flask\footnote{Flask: \url{https://flask.palletsprojects.com/}} service that exposes several REST endpoints. Both servers expose an endpoint that accepts a file via a POST request, runs it, and returns either a success or an error message depending on the execution result. The Flask service in the \ac{MEC} server also provides an endpoint that reports the current CPU usage, allowing the \ac{UE} to assess the server's status before offloading tasks. Table \ref{tab:flask_endpoints} shows a description of exposed methods.  

\begin{table}[htbp]
\centering
\caption{Servers' endpoints}
\label{tab:flask_endpoints}
%\resizebox{\columnwidth}{!}{%
\begin{tabular}{|c|c|c|c|c|}
\hline
\textbf{Server}               & \textbf{Feature} & \textbf{HTTP Method} & \textbf{Inputs} & \textbf{Outputs} \\ \hline
\multirow{2}{*}{\textbf{MEC}} & Run task         & POST                  & Task & OK / Error \\ \cline{2-5} 
               & Get load & GET  & \textbf{-}        & CPU usage        \\ \hline
\textbf{Cloud} & Run task & POST & Task & OK / Error \\ \hline
\end{tabular}%
%}
\end{table}

The servers' characteristics are:
\begin{itemize}
    \item \textbf{The \ac{MEC} server} runs an Ubuntu 20.04.3 LTS platform with a 4th Gen IntelCore i7 CPU 2.00 GHz and 8 GB of RAM. It is placed a few meters from the \ac{UE} and is physically connected to the \ac{BS} via an Ethernet cable, providing low-latency communication. The \ac{MEC} server can be subjected to stress using the \textit{stress-ng} tool\footnote{\textit{stress-ng}: \url{https://github.com/ColinIanKing/stress-ng}}, which simulates user-defined load conditions.

    \item \textbf{The \ac{CS}}, also running Ubuntu 20.04.3 LTS, is a 4th Gen Intel Core i7 CPU 2.50 GHz and 32 GB of RAM. The network conditions between the \ac{BS} and the \ac{CS} are altered through the addition of a \textbf{Network emulator}, which enables the emulation of latencies, jitter, packet loss, burst loss correlation, corruption, reordering, and packet duplication.
\end{itemize}

Finally, the TACLeBench suite task-set generator \cite{DeBock2018} is utilized to generate tasks, enabling the creation of executable tasks with varying characteristics, including different computational complexities and latency constraints. This tool generates representative executable tasks that reflect real-world applications, while allowing adjustments to task attributes such as computational load and timing constraints. Therefore, synthetic data is generated for each task to be offloaded, from which the latency requirement is derived similarly to the approach in \cite{Zhang2018}, where the delay requirement for each task $(\delta_i)$ is modeled as $C_i/\beta$, with $\beta$ uniformly distributed in $[9, 11]$. For this paper, tasks have been generated as per the configuration presented in Table \ref{tab:tasks_info}. 

\begin{table}[ht]
\caption{Tasks characteristics}
\label{tab:tasks_info}
\centering 
\begin{tabular}{c|c|c|}
\cline{2-3}
                       & \textbf{Parameter}     & \textbf{Value} \\ \hline
\multicolumn{1}{|c|}{\multirow{9}{*}{\textbf{TaskSet}}}   & Min global utilization & 3.9 \\ \cline{2-3} 
\multicolumn{1}{|c|}{} & Max global utilization & 3.9            \\ \cline{2-3} 
\multicolumn{1}{|c|}{} & Utilization step       & 0.1            \\ \cline{2-3} 
\multicolumn{1}{|c|}{} & Number of Tasks        & 500            \\ \cline{2-3} 
\multicolumn{1}{|c|}{} & Number of Tasksets     & 1              \\ \cline{2-3} 
\multicolumn{1}{|c|}{} & Min period             & 1e5            \\ \cline{2-3} 
\multicolumn{1}{|c|}{} & Max period             & 1e8            \\ \cline{2-3} 
\multicolumn{1}{|c|}{} & Period step            & 1000           \\ \cline{2-3} 
\multicolumn{1}{|c|}{} & Seed                   & 50             \\ \hline
\multicolumn{1}{|c|}{\multirow{3}{*}{\textbf{Target HW}}} & Architecture           & x86 \\ \cline{2-3} 
\multicolumn{1}{|c|}{} & CPU Speed (MHz)        & 1200.048       \\ \cline{2-3} 
\multicolumn{1}{|c|}{} & Cores                  & 4              \\ \hline
\end{tabular}%
\end{table}

\subsection{Training process}

This system implements the capability of distributed training for \ac{DRL} agents (either \ac{AC} or \ac{DQN}) using the \ac{MQTT} protocol as a lightweight communication layer between clients or \acp{UE} and a training server located in the \ac{MEC} server.

Therefore, each \ac{UE} runs a \ac{DRL} agent (created using \textit{Tensorflow}\footnote{Tensorflow: \url{https://www.tensorflow.org/}} and \textit{Keras}\footnote{Keras: \url{https://keras.io/}}) that interacts with the environment and takes actions on-device. Regarding the training process, the model can be trained locally or collect and send experience tuples (state, action, reward, and next state) to the \ac{MEC} server via \ac{MQTT}. When the server sends back updated \ac{NN} weights, the client updates its local model immediately, allowing it to choose actions for the new incoming tasks using the updated model.

The \ac{MEC} server acts as a global learner that subscribes to \ac{MQTT} topics from all \acp{UE}, receives serialized experiences as \textit{protobuf} messages, updates the corresponding model using the received data, and publishes the updated model weights back to the corresponding \ac{UE}. This way, the \ac{MEC} server can manage multiple agents concurrently, identified by device ID $D$ and model type $\Pi$.

Considering that some \acp{UE} include a GPU, \textit{TensorFlow} can automatically leverage GPU acceleration if available. For GPU-equipped devices, such as the \textit{Jetson AGX Orin}, GPU usage can be explicitly enabled or disabled through the device's configuration. The \ac{MEC} server does not include a GPU; therefore, even if the \ac{UE} supports GPU acceleration, training performed on the server will not utilize it. The pseudo-code of the Training process is explained in Algorithm~\ref{alg:mqtt_generic}. 

\begin{algorithm}[H]
\caption{Distributed Reinforcement Learning Framework}
\label{alg:mqtt_generic}
\begin{algorithmic}[1]
\Require \textit{Model type $\Pi$}, \textit{Device ID $D$}, \textit{Broker address}, \textit{Training mode}

%\Statex
%\State \textbf{Initialization:}
\State Initialize \textit{Agent $A$} using \textit{Model type $\Pi$} with parameters $\theta_A$
% \If{$\text{\textit{Training mode = }remote}$}
%     \State Connect to MQTT broker $B$
%     \State Subscribe to topic \texttt{server/D/weights/M/v1}
% \Else
%     \State Initialize local training process
% \EndIf

%\Statex
\While{training not finished}
    \State Observe current state $s_i$
    \State Select action $\alpha_i \sim \pi(s_i; \theta_A)$
    \State Execute $\alpha_i$, observe reward $r_i$, next state $s_{i+1}$

    \If{$\textit{Training mode = }remote$}
        \State Package $(s_i, \alpha_i, r_i, s_{i+1})$ as \texttt{Experience} message
        \State Publish to topic \texttt{client/D/experience/M/v1}
        %\State Subscribe to \texttt{client/+/experience/+/v1}
        \For{each received experience $(s_i, \alpha_i, r_i, s_{i+1})$}
            \State Identify agent type $M$
            \State Perform one training step on $\theta$
            \State Broadcast $\theta$ via \texttt{Weights} message
        \EndFor
    \Else
        \If{\text{GPU is chosen}}
            \State Train agent locally using GPU and experience $(s_i, \alpha_i, r_i, s_{i+1})$
        \Else
            \State Train agent locally using CPU and experience $(s_i, \alpha_i, r_i, s_{i+1})$
        \EndIf
    \EndIf

    \If{received new \texttt{Weights} message}
        \State Update local parameters $\theta_A \gets \theta_{server}$
    \EndIf
\EndWhile

% \Statex
% \Statex \textbf{Server process (only if remote = True):}
% \State Subscribe to \texttt{client/+/experience/+/v1}
% \For{each received experience $(s_i, \alpha_i, r_i, s_{i+1})$}
%     \State Identify agent type $M$
%     \State Perform one training step on $\theta$
%     \State If update interval reached, broadcast $\theta$ via \texttt{Weights} message
% \EndFor

%\Statex
%\State \textbf{Output:} Trained parameters $\theta_A$
\end{algorithmic}
\end{algorithm}

There, the topic structure \textit{client/D/experience/M/v1} encodes the device ID \textit{D} and model type \textit{M}, allowing the server to route experiences to the correct learner instance. The version suffix \textit{v1} provides namespace isolation, enabling future model revisions.

Each device runs its own version of \textit{TensorFlow}, tailored to its Hardware. \textit{reComputer J1010} and \textit{Raspberry Pi 4} use \textit{Tensorflow} version 2.14, while \textit{Jetson AGX Orin} uses \textit{Tensorflow} version 2.16.

\begin{figure}[hbt!]
    \center
    \includegraphics[width=0.8\columnwidth]{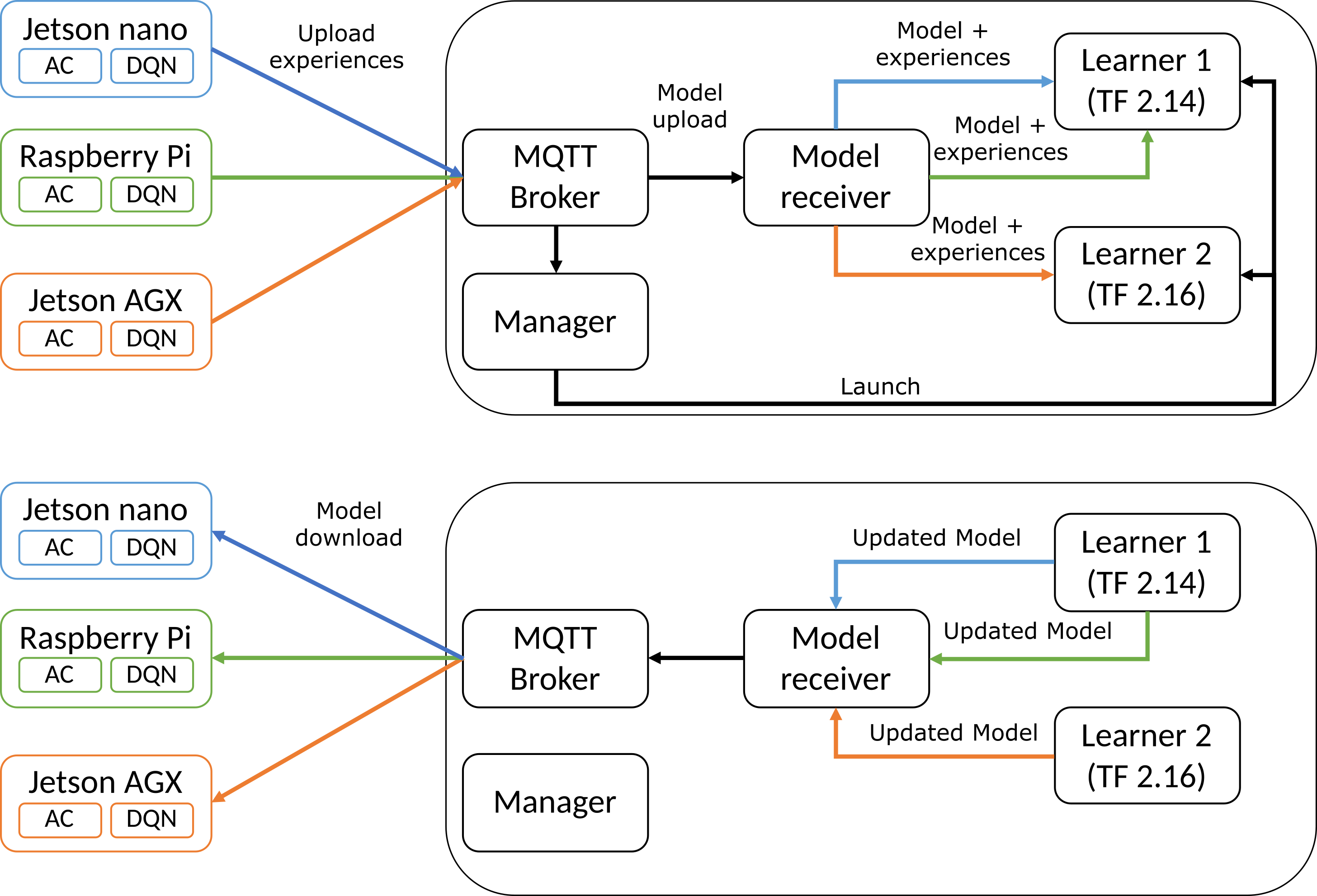}
    \caption{MQTT Process}
    \label{fig:mqtt_process}
\end{figure}

A workflow schema is presented in Figure \ref{fig:mqtt_process}. The devices publish their experience data and upload their own \ac{DRL} models. The \ac{MQTT} broker runs on the \ac{MEC} server and listens on all network interfaces, receiving experiences through the Server Learner and managing incoming models through the Model receiver. The Server Model Receiver subscribes to upload topics from devices and collects and assembles model chunks. Regarding the server learner, each Server Learner process is launched by the Server Manager with the correct virtual environment for its device type. This way, it subscribes to its assigned experience topic, uses the received experiences to train a \ac{DRL} agent locally on the server, and publishes updated model weights back to devices via \ac{MQTT}. This design allows each learner to specialize per device and model type while sharing compute resources centrally. 

% \subsection{Optimization}
% \begin{itemize}
%     \item Mixed float16 float32: Error 
%     \item QAT: ValueError: Unable to clone model. This generally happens if you used custom Keras layers or objects in your model. Please specify them via `quantize scope` for your calls to `quantize model` and `quantize apply`. [Layer PruneLowMagnitude supplied to wrapper is not a supported layer type. Please ensure wrapped layer is a valid Keras layer.].
%     \item Gradual pruning
% \end{itemize}

%Avoid batch training: Es mejor no hacer batch training o batches muy pequeños para que sea más eficiente

\section{Results}
\label{sec:results}

This Section presents a comprehensive evaluation of on-device and remote \ac{DRL} training across heterogeneous edge platforms. The results are analyzed focusing on training time, energy consumption, and accumulated reward, with the goal of assessing the practical feasibility and efficiency of on-device learning in realistic edge environments. 

We evaluate on-device \ac{DRL} in continual learning settings, so each device learns on the fly for every incoming task. This way, for each task, the device not only runs inference using the \ac{DRL} policy but also updates the deployed agent, performing task-specific training rather than batch training. 

An initial characterization of the devices was performed to compare their peak computational capabilities, which was conducted using openBLAS\footnote{OpenBLAS: \url{http://www.openmathlib.org/OpenBLAS/}} and cuBLAS\footnote{cuBLAS: \url{https://developer.nvidia.com/cublas}}, and results are presented in Table \ref{tab:device_flops}. \ac{DRL} involves repeated execution of forward and backward passes during training, which are computationally intensive due to the underlying matrix operations. To assess the theoretical performance of these operations on the evaluated platforms, we measure \ac{SGEMM} and \ac{DGEMM} on both CPU and GPU. \ac{SGEMM} represents the dominant operation in typical neural-network inference and training, while \ac{DGEMM} provides insight into double-precision performance for tasks requiring higher numerical accuracy. The benchmarks also indicate that \ac{NN} training will benefit from GPU acceleration on a given device, regardless of whether inference is likely to be CPU-bound. It is worth noting that the \textit{Raspberry Pi} and r\textit{eComputer J1010 / Jetson Nano} support up to 4 CPU threads, whereas the \textit{Jetson AGX Orin} supports up to 8 threads. Regarding GPU capabilities, the \textit{Jetson AGX Orin} 64GB features 2,048 CUDA cores\footnote{\textit{Jetson AGX Orin} datasheet: \url{https://nvidia.com/content/dam/en-zz/Solutions/gtcf21/jetson-orin/nvidia-jetson-agx-orin-technical-brief.pdf}}, while the \textit{reComputer J1010} includes 128 CUDA cores\footnote{\textit{reComputer J1010} datasheet: \url{https://files.seeedstudio.com/wiki/reComputer/reComputer-J1010-datasheet.pdf}}.

\begin{table}[htbp]
\caption{Operations per second for each device}
\label{tab:device_flops}
\resizebox{\columnwidth}{!}{%
\centering
\begin{tabular}{|ccc|cc|cc|}
\hline
\multicolumn{3}{|c|}{\textbf{Device}} &
  \multicolumn{2}{c|}{\textbf{SGEMM (GFLOPS)}} &
  \multicolumn{2}{c|}{\textbf{DGEMM (GFLOPS)}} \\ \hline
\multicolumn{1}{|c|}{\textbf{Name}} &
  \multicolumn{1}{c|}{\textbf{HW}} &
  \textbf{\begin{tabular}[c]{@{}c@{}}CPU /\\ GPU\end{tabular}} &
  \multicolumn{1}{c|}{\textbf{\begin{tabular}[c]{@{}c@{}}Mono\\ Thread\end{tabular}}} &
  \textbf{\begin{tabular}[c]{@{}c@{}}Multi\\ Thread\end{tabular}} &
  \multicolumn{1}{c|}{\textbf{\begin{tabular}[c]{@{}c@{}}Mono\\ Thread\end{tabular}}} &
  \textbf{\begin{tabular}[c]{@{}c@{}}Multi\\ Thread\end{tabular}} \\ \hline
\multicolumn{1}{|c|}{\multirow{2}{*}{\begin{tabular}[c]{@{}c@{}}Jetson\\ AGX Orin\end{tabular}}} &
  \multicolumn{1}{c|}{\multirow{2}{*}{\begin{tabular}[c]{@{}c@{}}12x A78AE\\ 2.2 GHz\end{tabular}}} &
  CPU &
  \multicolumn{1}{c|}{24.03} &
  108.93 &
  \multicolumn{1}{c|}{12.36} &
  70.19 \\ \cline{3-7} 
\multicolumn{1}{|c|}{} &
  \multicolumn{1}{c|}{} &
  GPU &
  \multicolumn{1}{c|}{-} &
  435.61 &
  \multicolumn{1}{c|}{-} &
  14.66 \\ \hline
\multicolumn{1}{|c|}{\multirow{2}{*}{\begin{tabular}[c]{@{}c@{}}reComputer J1010\\ - Jetson nano\end{tabular}}} &
  \multicolumn{1}{c|}{\multirow{2}{*}{\begin{tabular}[c]{@{}c@{}}4xA57\\ 1.4 GHz\end{tabular}}} &
  CPU &
  \multicolumn{1}{c|}{9.11} &
  29.50 &
  \multicolumn{1}{c|}{1.60} &
  6.10 \\ \cline{3-7} 
\multicolumn{1}{|c|}{} &
  \multicolumn{1}{c|}{} &
  GPU &
  \multicolumn{1}{c|}{-} &
  28.09 &
  \multicolumn{1}{c|}{-} &
  6.31 \\ \hline
\multicolumn{1}{|c|}{Raspberry Pi 4b} &
  \multicolumn{1}{c|}{\begin{tabular}[c]{@{}c@{}}4xA72\\ 1.5 GHz\end{tabular}} &
  CPU &
  \multicolumn{1}{c|}{8.54} &
  20.06 &
  \multicolumn{1}{c|}{4.56} &
  12.25 \\ \hline
\end{tabular}%
}
\end{table}

We use a task set composed of diverse synthetic tasks generated by the open-source framework for benchmarking and \ac{WCET} research described in Section \ref{sec:setup}. %COBRA \cite{cobra}, an open-source framework for benchmarking and \ac{WCET} research. Additionally, several types of benchmarks are included, namely kernel benchmarks to implement small kernel functions, sequential benchmarks to allow for large function blocks such as encoders and decoders commonly used in embedded systems, parallel benchmarks, and other real-world operation-specific benchmarks. 

Finally, regarding the tests, different scenarios can be configured to evaluate performance under various conditions. These configuration options are shown in Table \ref{tab:config-options}.

\begin{table}[ht]
\centering
\caption{Configurable parameters and options with device-specific constraints}
\label{tab:config-options}
\resizebox{\columnwidth}{!}{%
\begin{tabular}{cc}
\hline
\textbf{Parameter}       & \textbf{Options} \\ \hline
Device                   & Raspberry Pi $\vert\vert$ Jetson Nano $\vert\vert$ Jetson AGX Orin \\
Algorithm                & \ac{AC} $\vert\vert$ \ac{DQN} \\
Training Location        & Local $\vert\vert$ \ac{MEC} \\
MEC Stress               & No stress $\vert\vert$ Stress \\
Mode                     & CPU $\vert\vert$ GPU (not available on Raspberry Pi) \\
Multi-UE                 & No multi-UE $\vert\vert$  Multi-UE\\ \hline
\end{tabular}
}
\end{table}

Therefore, we test the system under two different scenarios:
\begin{itemize}
    \item \textbf{Scenario 1: Ideal conditions.} In this scenario, there is no stress nor virtual \ac{UE} interacting with the network.
    \item \textbf{Scenario 2: Concurrent \acp{UE}.} In this scenario, we emulate 50 concurrent \acp{UE}, each configured to generate traffic toward the \ac{MEC} server using \textit{iperf}\footnote{Iperf: \url{https://iperf.fr/}}. The aggregate load corresponds to 50 \acp{UE} transmitting at 10 Mbps each. This setup also includes a stress on the \ac{MEC} server that fluctuates between 80\% and 100\% of its processing capacity, following a predefined distribution, allowing us to evaluate performance under heavy multi-user conditions and assess scalability in real-world deployments.
\end{itemize}

% \begin{table}[]
% \caption{Configurable parameters and options with device-specific constraints}
% \label{tab:config-options}
% \resizebox{\columnwidth}{!}{%
% \begin{tabular}{|c|cccc|}
% \hline
% \textbf{Parameter}  & \multicolumn{4}{c|}{\textbf{Options}}                                             \\ \hline
% Device & \multicolumn{1}{c|}{Raspberry Pi} & \multicolumn{2}{c|}{\begin{tabular}[c]{@{}c@{}}Recomputer   J10\\ – Jetson nano\end{tabular}} & Jetson AGX   Orin \\ \hline
% Algorithm           & \multicolumn{2}{c|}{AC}          & \multicolumn{2}{c|}{DQN}                       \\ \hline
% Training   Location & \multicolumn{2}{c|}{Local}       & \multicolumn{2}{c|}{MEC}                       \\ \hline
% MEC Stress & \multicolumn{2}{c|}{No stress}                               & \multicolumn{2}{c|}{Stress   \textgreater{}80\%} \\ \hline
% Mode                & \multicolumn{2}{c|}{CPU}         & \multicolumn{2}{c|}{GPU (not in Raspberry Pi)} \\ \hline
% Multi-UE            & \multicolumn{2}{c|}{No Multi-UE} & \multicolumn{2}{c|}{UEs x 10 M BW iperf}       \\ \hline
% \end{tabular}%
% }
% \end{table}

\begin{figure}[htbp]
    \centering
    \includegraphics[width=0.8\linewidth]{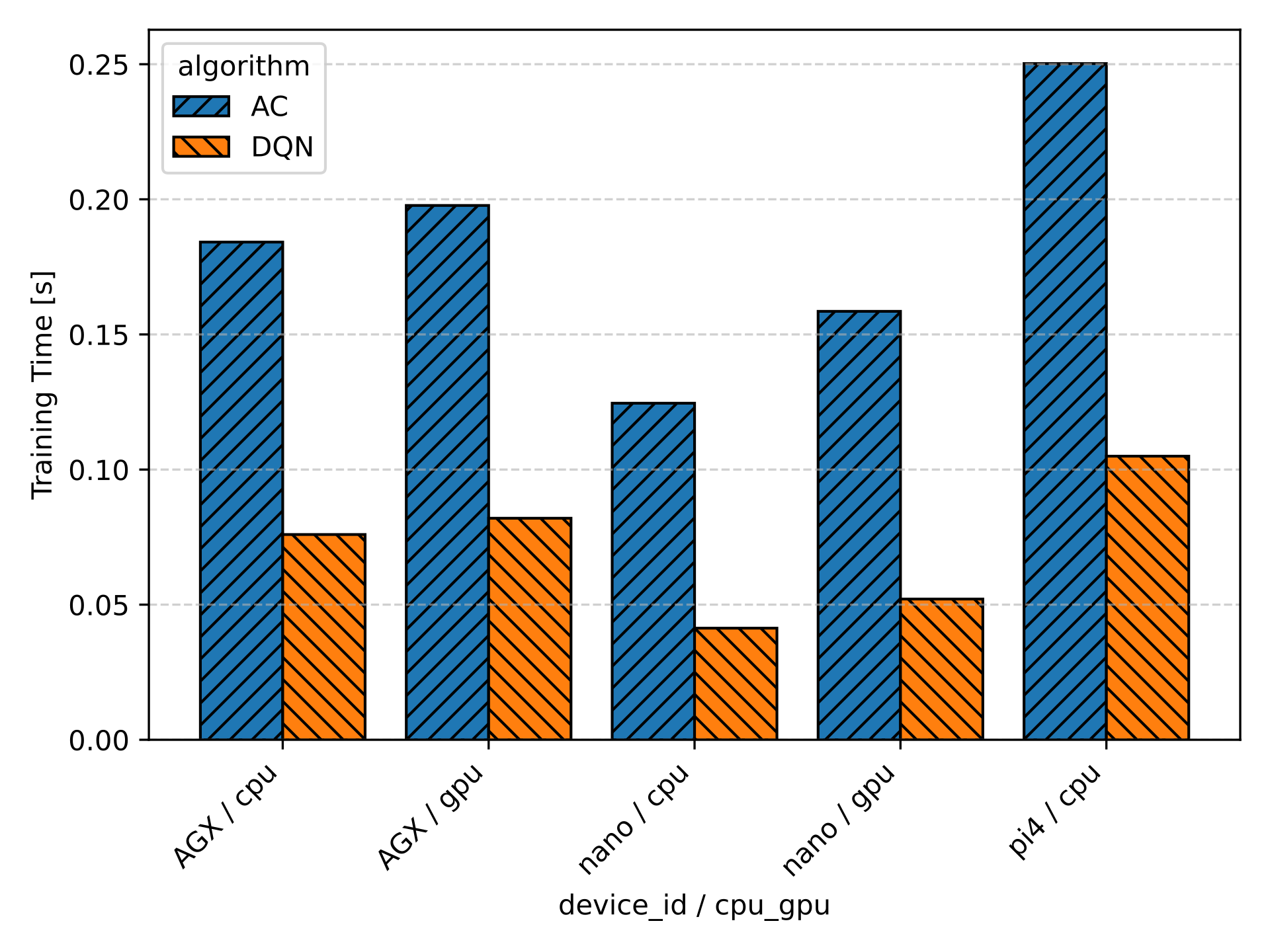}
    \caption{Average local training time for each algorithm in each device in Scenario 1}
    \label{fig:dev_train_time}
\end{figure}

First, we run both \ac{DRL} agents locally in Scenario 1 (under ideal conditions), as shown in Figure~\ref{fig:dev_train_time}. Across all devices, \ac{AC} consistently exhibits longer training times compared to \ac{DQN}, but the behavior is similar across all devices using both algorithms. Furthermore, GPU acceleration does not improve \ac{DRL} training performance in the evaluated edge scenarios, as the additional data transfers and kernel launch overheads lead to higher energy consumption and longer convergence times compared to CPU-based training. Additionally, the lack of large batches reduces parallelization efficiency, resulting in underutilization of GPU cores and increased latency per update. %. Considering that these experiments involve sequential training for each incoming task without batching, the overhead of frequent model updates and synchronization amplifies the computational demands, which has more impact on \ac{AC}. Furthermore, regarding GPU usage,
%, as the lack of large batches reduces parallelization efficiency, resulting in underutilization of GPU cores and increased latency per update.
In contrast, CPUs handle smaller, frequent updates more effectively. \textit{Raspberry Pi} shows the worst training times, while \textit{reComputer J1010} shows the best ones.

\begin{figure}[htbp]
    \centering
    \includegraphics[width=0.8\linewidth]{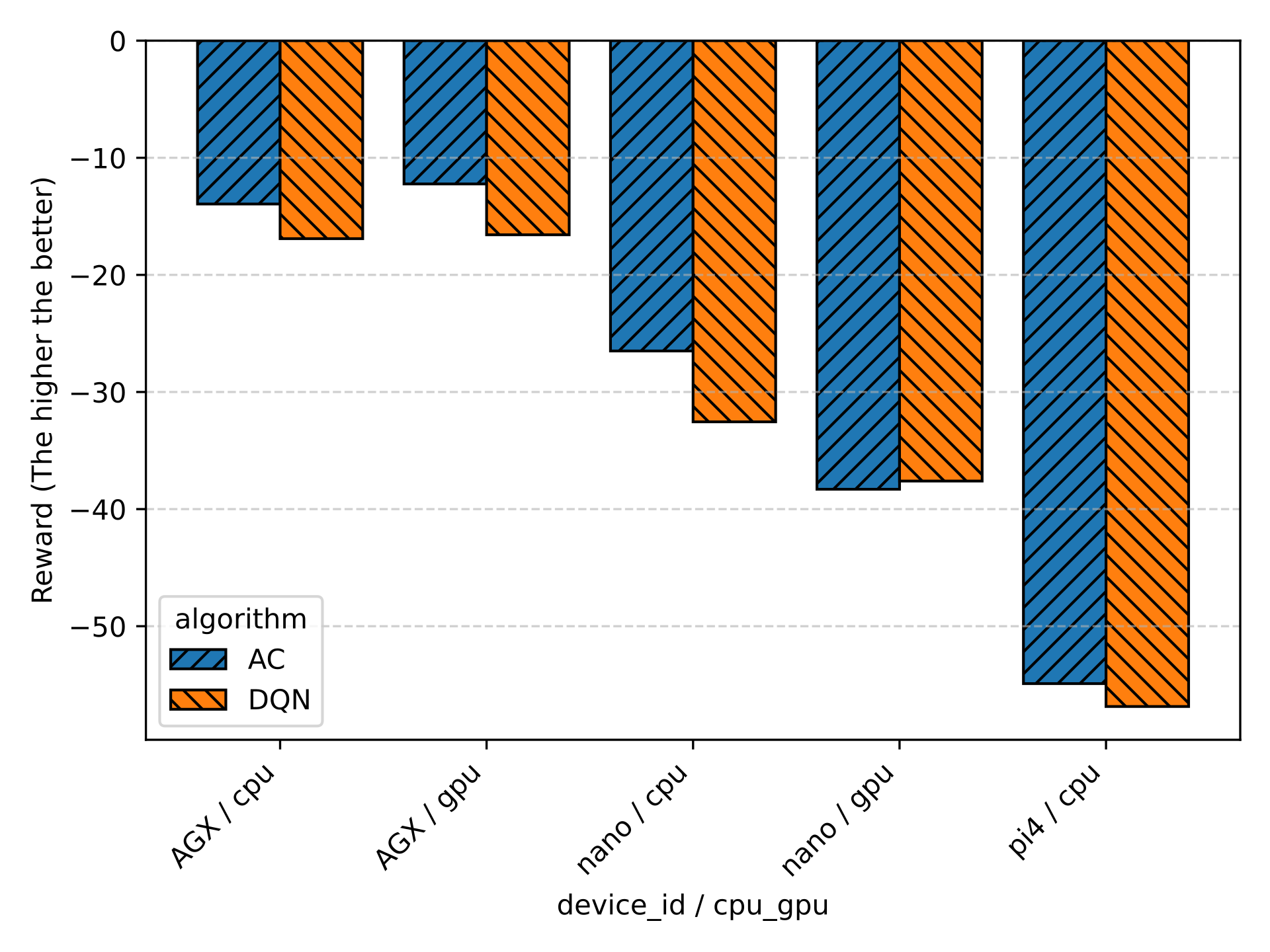}
    \caption{Average reward for each algorithm in each device in Scenario 1}
    \label{fig:dev_reward}
\end{figure}

In the same scenario, we assess the performance of both \ac{DRL} algorithms across every edge device with local training. The best results are observed on the \textit{Jetson AGX Orin}, as its larger computational capabilities allow the algorithms to run more tasks locally, resulting in lower latencies. In contrast, the reward significantly degrades on resource-constrained platforms such as the \textit{Raspberry Pi} and \textit{reComputer J1010}. Their limited processing power leads to longer delays when running locally, and they often decide to offload more frequently, for both \ac{AC} and \ac{DQN}. Figure~\ref{fig:dev_reward} shows the performance of the algorithms in terms of the reward, as defined in Section \ref{sec:prob_form}.

\begin{figure}[htbp]
    \centering
    \includegraphics[width=0.8\linewidth]{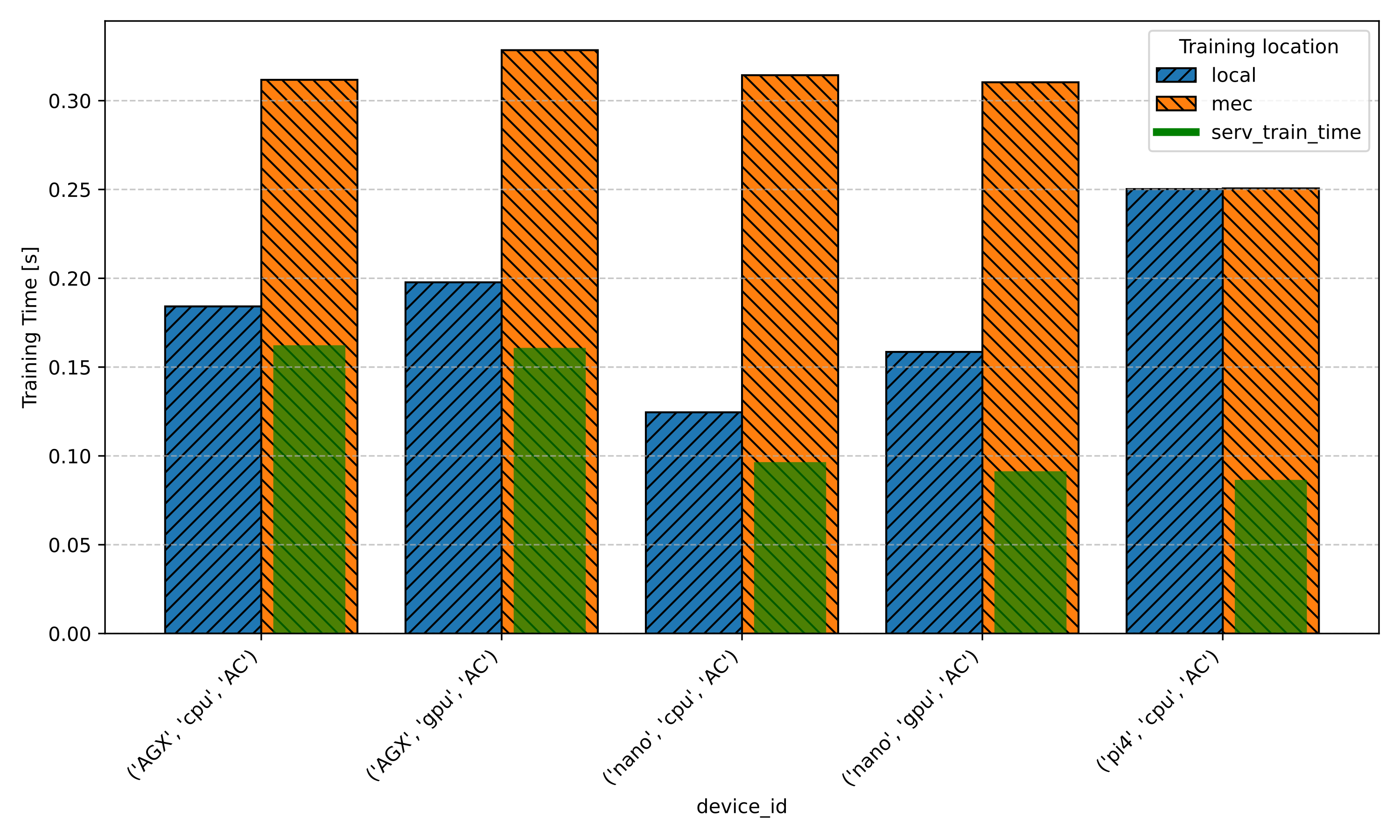}
    \caption{Average \ac{AC} training time locally and remotely for each device in Scenario 1}
    \label{fig:ac_local_mec_time}
\end{figure}

Then, in Figure~\ref{fig:ac_local_mec_time}, we compare the training time when executed locally on the device versus remotely on the \ac{MEC} server in Scenario 1 using the \ac{AC} algorithm. Overall, local training is consistently faster due to the absence of communication overhead. When training on the \ac{MEC}, the duration is relatively similar across all devices, with minor variations caused by network conditions and modem behavior. %\textcolor{red}{Notably, the \textit{Raspberry Pi} achieves faster communication through the Quectel modem compared to Jetson-based devices (\textit{AGX Orin} and \textit{reComputer J1010}). This difference is primarily due to the Raspberry Pi running Raspberry OS, which includes well-tested drivers for USB modems such as QMI. In contrast, Jetson devices rely on customized kernels that lack full compatibility, resulting in slightly degraded performance.} 
It is essential to note that the \ac{MEC} training time encompasses the entire process, including sending the model, performing training (as shown in the green column), receiving the updated parameters, and reassembling the model.

\begin{figure}[htbp]
    \centering
    \includegraphics[width=0.8\linewidth]{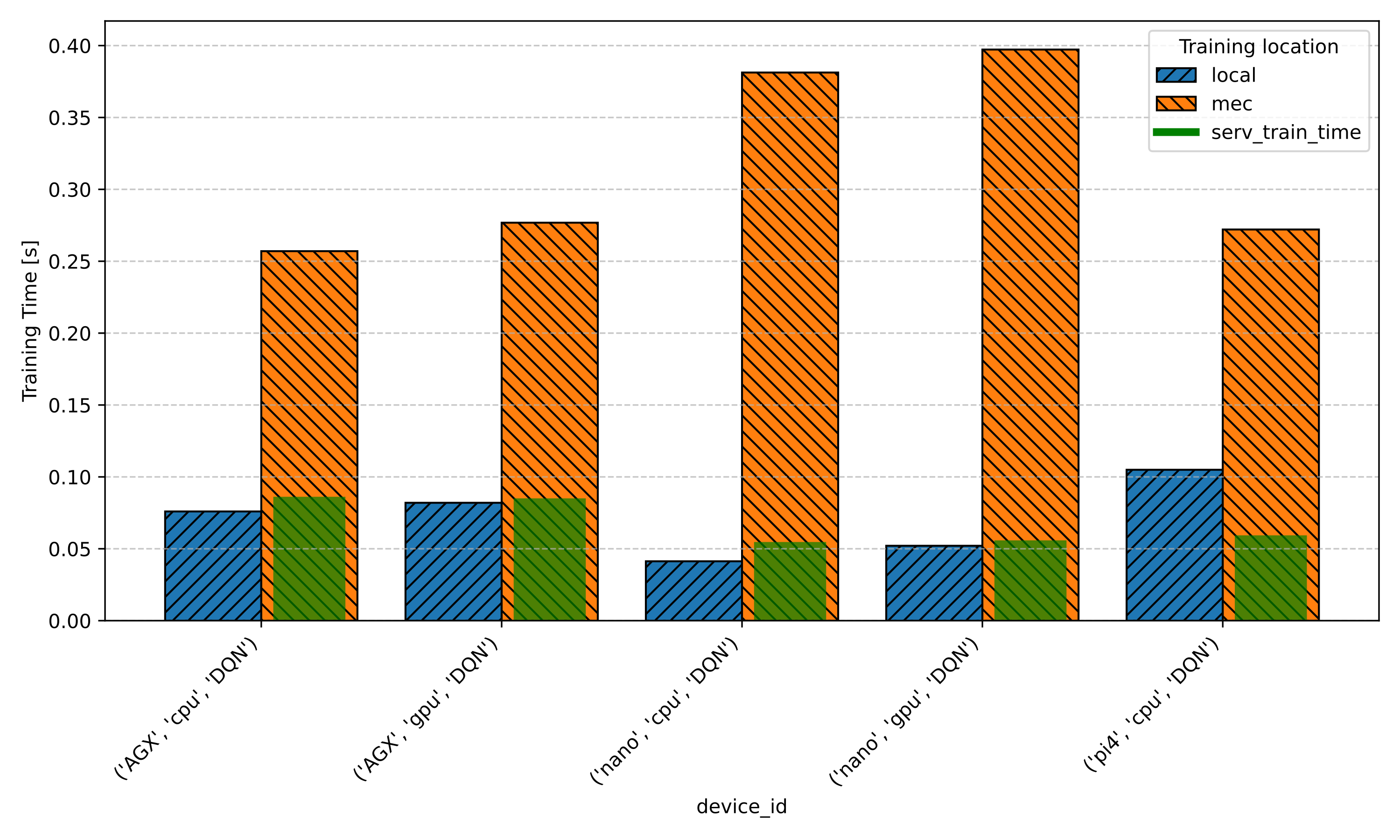}
    \caption{Average \ac{DQN} training time locally and remotely for each device in Scenario 1}
    \label{fig:dqn_local_mec_time}
\end{figure}

A similar pattern is observed in Figure~\ref{fig:dqn_local_mec_time}, which shows the training time when using the \ac{DQN} algorithm across all devices, training locally and remotely on the \ac{MEC} server. Training on the \ac{MEC} server, which includes transmission, training, and reception times, remains consistent across all devices, with times ranging from 0.25 to 0.35 seconds. However, the training process on the server is slightly longer using \textit{Jetson AGX Orin} due to the different TensorFlow version. In contrast, local training is significantly faster and notably shorter than using \ac{AC}, reflecting the lower computational complexity of \ac{DQN}. This algorithm is less computationally demanding to train, which explains the substantial difference in training times between local and remote settings. These differences highlight the impact of algorithmic complexity on deployment strategies.

\begin{figure}[htbp]
    \centering
    \includegraphics[width=0.8\linewidth]{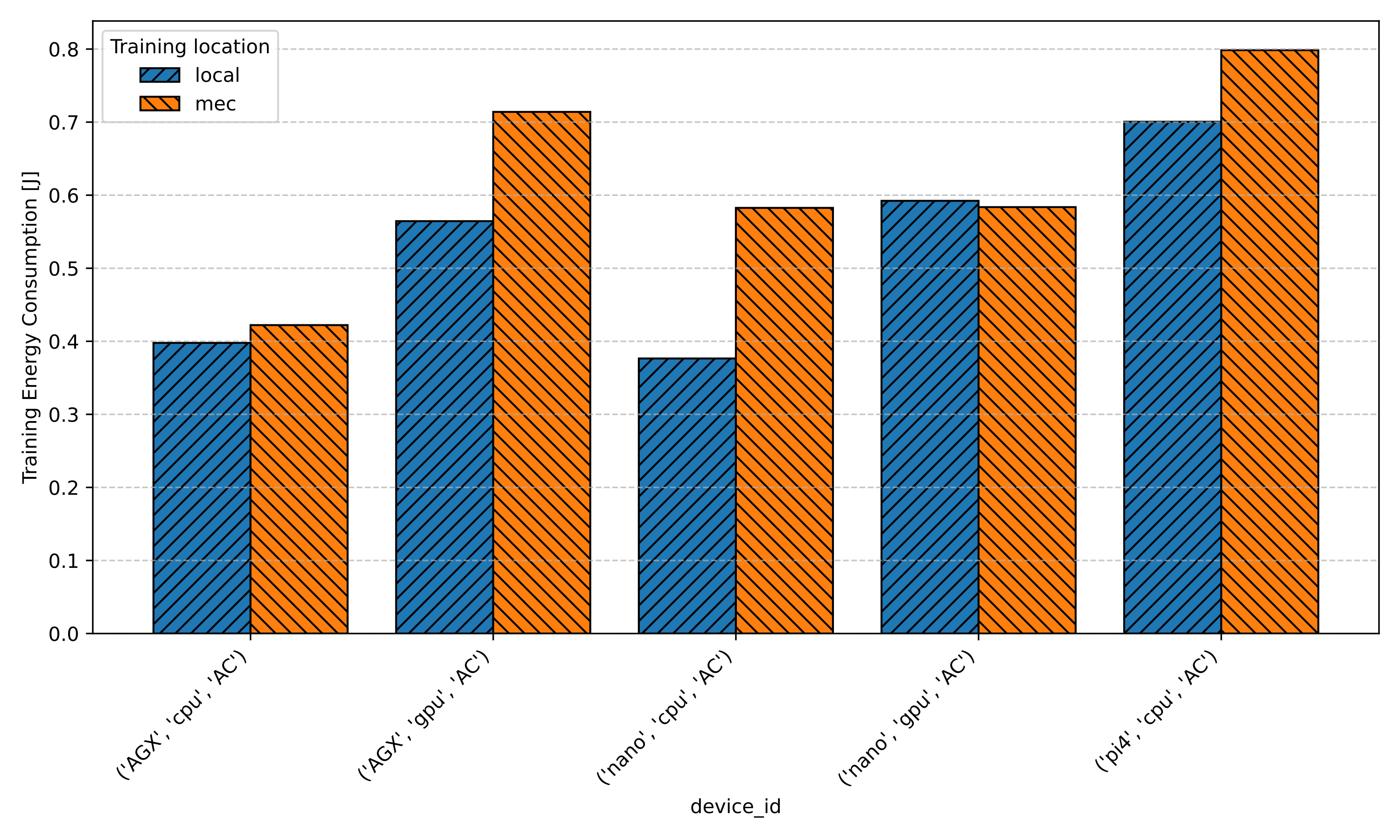}
    \caption{Average \ac{AC} training energy locally and remotely for each device in Scenario 1}
    \label{fig:ac_local_mec_energy}
\end{figure}

We analyze the energy consumption on-device due to training the \ac{AC} agent in Scenario 1, both locally and remotely on the \ac{MEC} server. Since energy usage is closely related to the time spent during training, offloading to the \ac{MEC} results in higher overall power consumption due to additional communication overhead. This includes idle periods while waiting for data transfer and model updates. Figure~\ref{fig:ac_local_mec_energy} illustrates the energy consumption across \textit{reComputer J1010}, \textit{Jetson AGX Orin}, and \textit{Raspberry Pi}, highlighting that remote training consistently incurs greater energy costs compared to local execution.

% Energy correlated with time, but power consumption is higher when running the model locally \ref{tab:ac_dqn_power_raspi}. \textcolor{red}{ESTO ESTÁ MAL CALCULADO. Dejar claro qeu es el consumo en el dispositivo. Que no piensen que estoy midiendo el consumo energético del MEC. Cambiar tambien el caption}

% \begin{table}[htbp]
%     \centering
%     \caption{Power consumption of training in Raspberry Pi for each algorithm locally and remotely}
%     \begin{tabular}{c|c|c|}
%     \cline{2-3}
%                                                 & \ac{AC} & \ac{DQN} \\ \hline
%     \multicolumn{1}{|c|}{local}                 & $0.35\text{W}$       & $0.50\text{W}$        \\ \hline
%     \multicolumn{1}{|c|}{\ac{MEC}} & $0.31\text{W}$       & $0.30\text{W}$        \\ \hline
%     \end{tabular}
%     \label{tab:ac_dqn_power_raspi}
% \end{table}

\begin{figure}[htbp]
    \centering
    \includegraphics[width=0.8\linewidth]{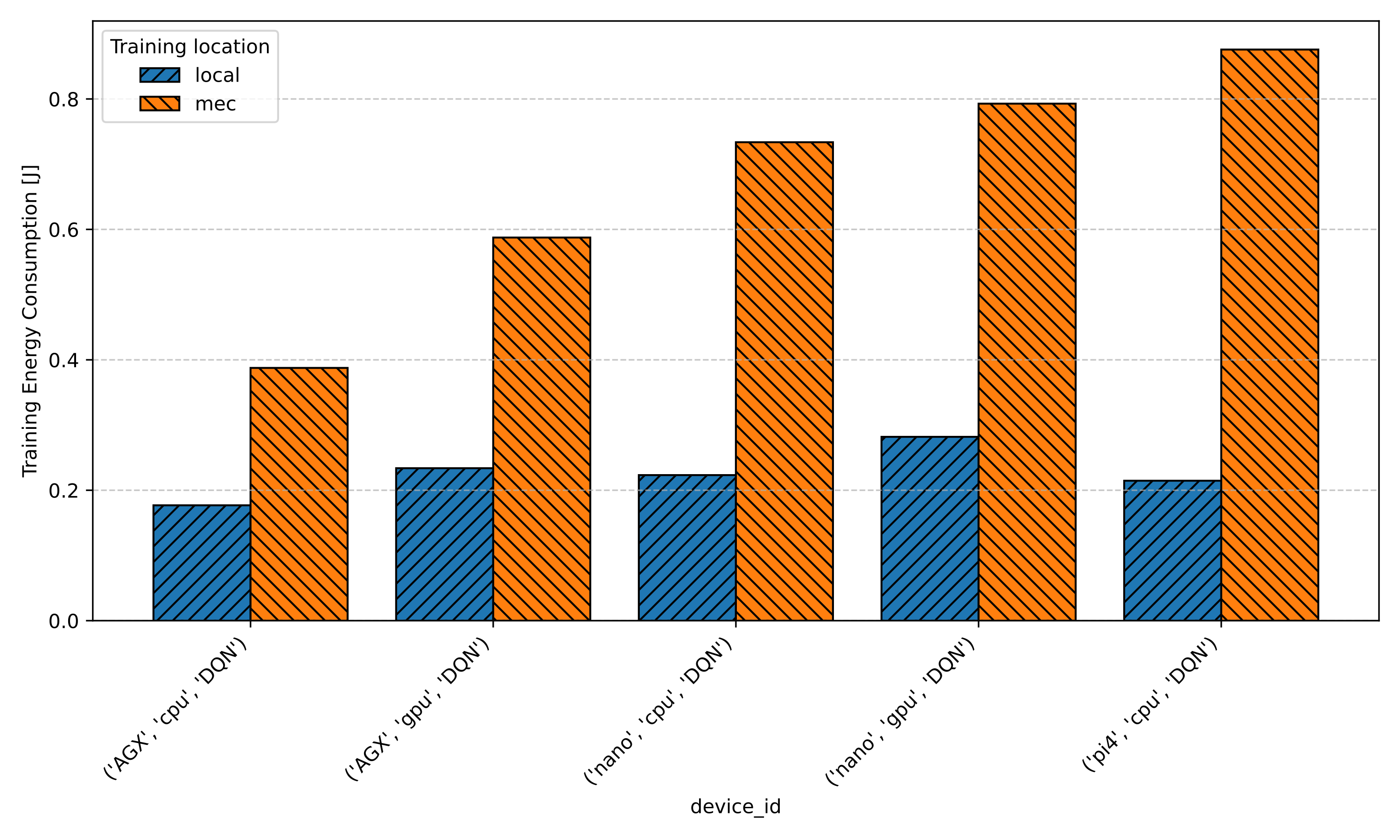}
    \caption{Average \ac{DQN} training energy locally and remotely for each device}
    \label{fig:dqn_local_mec_energy}
\end{figure}

The behavior when using \ac{DQN} is similar to that observed with \ac{AC}, as seen in Figure~\ref{fig:dqn_local_mec_energy}. However, the difference between local and remote training times is even more pronounced for \ac{DQN}, as local training is substantially faster due to its lower computational complexity. This highlights that offloading \ac{DQN} training to the \ac{MEC} introduces a disproportionately large overhead compared to local execution, making local training far more efficient for this algorithm.

\begin{figure}[htbp]
    \centering
    \includegraphics[width=\linewidth]{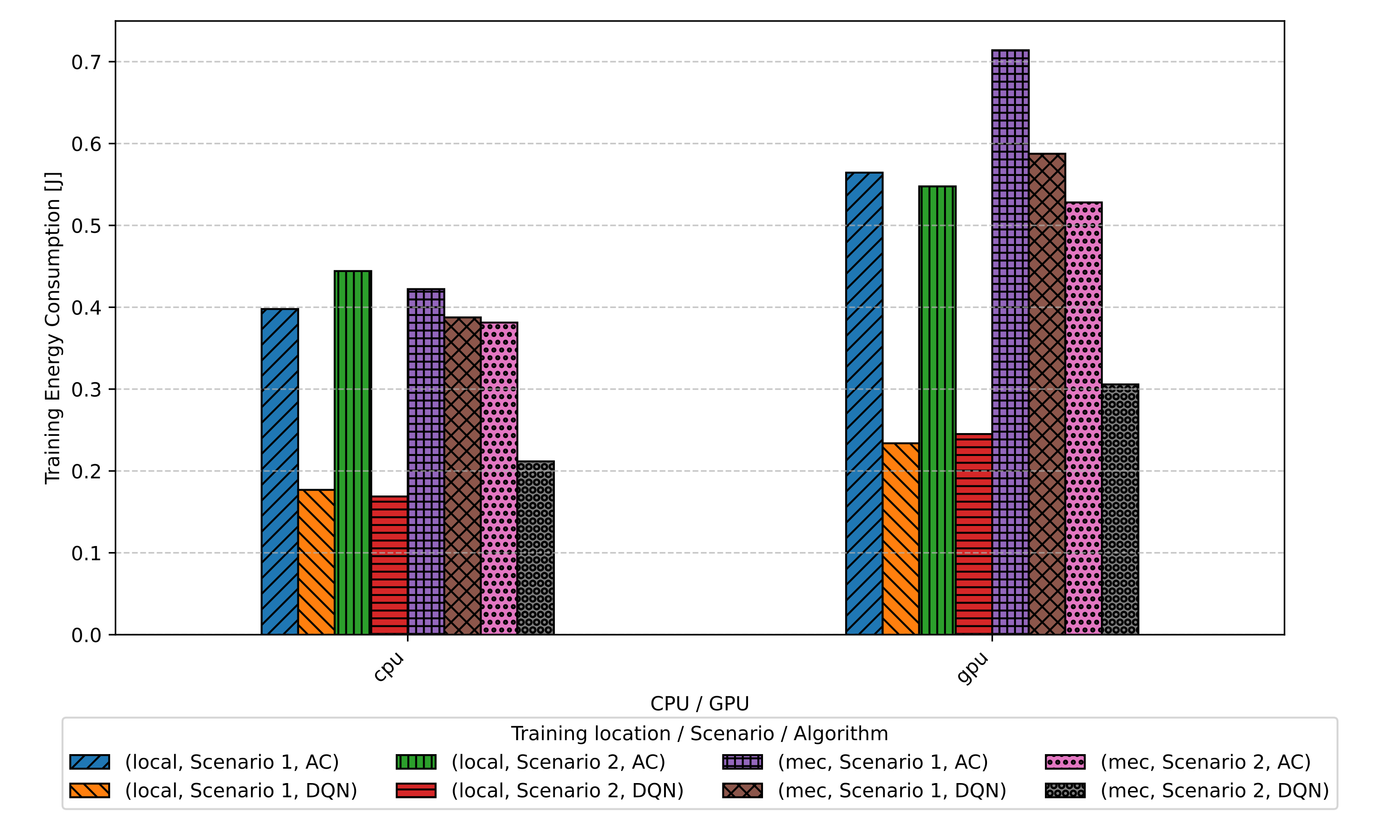}
    \caption{Consumed energy in AGX for each scenario}
    \label{fig:energy_comparison_agx}
\end{figure}

Finally, in general, energy consumption is higher when using the GPU, so its use is inappropriate in this context, as explained earlier. Figure~\ref{fig:energy_comparison_agx} shows the energy consumption during training for both Scenarios 1 and 2 using \ac{AC} and \ac{DQN} on the \textit{Jetson AGX Orin}. As observed, every test consumes more energy when the GPU is employed, as training times are also slightly longer compared to CPU execution. Therefore, GPU usage does not provide any advantage for these workloads and is not recommended.

\begin{figure}[htbp]
    \centering
    \includegraphics[width=\linewidth]{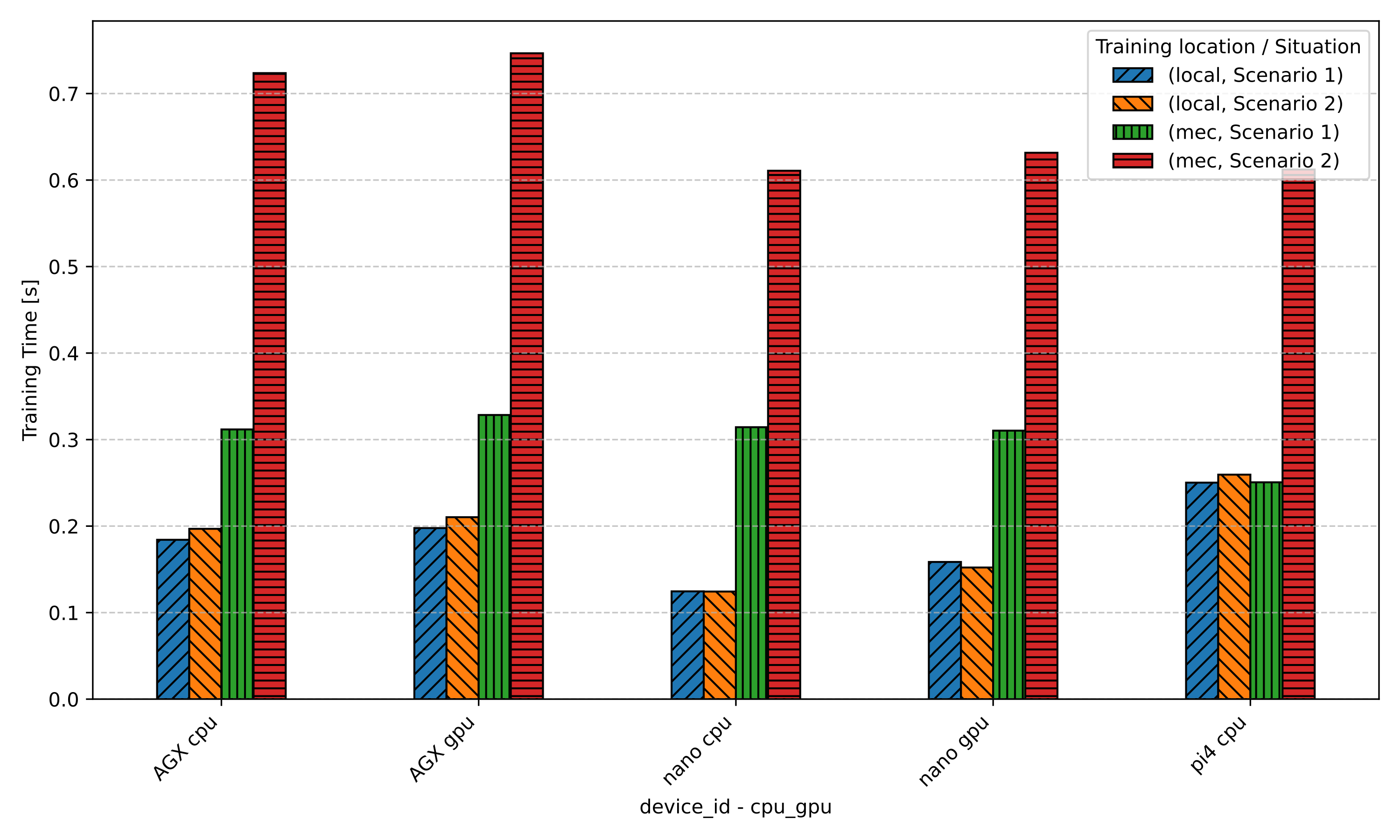}
    \caption{Comparison of training time of \ac{AC} in Scenarios 1 and 2}
    \label{fig:ac_time_compare_multiue}
\end{figure}

We then analyze the behavior of the \ac{AC} algorithm in Scenarios 1 and 2. In Scenario 2, the presence of multiple concurrent \acp{UE} introduces perturbations in both the communication channel and the \ac{MEC} server due to the additional traffic and computational demand. These conditions can significantly degrade remote training performance, as evidenced in Figure~\ref{fig:ac_time_compare_multiue}. When training is offloaded to the \ac{MEC} in Scenario 2, the training time increases substantially compared to Scenario 1, highlighting the impact of network congestion and resource contention on learning efficiency. The only situation where training on the \ac{MEC} server can be comparable to local training is when the \textit{Raspberry Pi} is used in Scenario 1.

\begin{figure}[htbp]
    \centering
    \includegraphics[width=\linewidth]{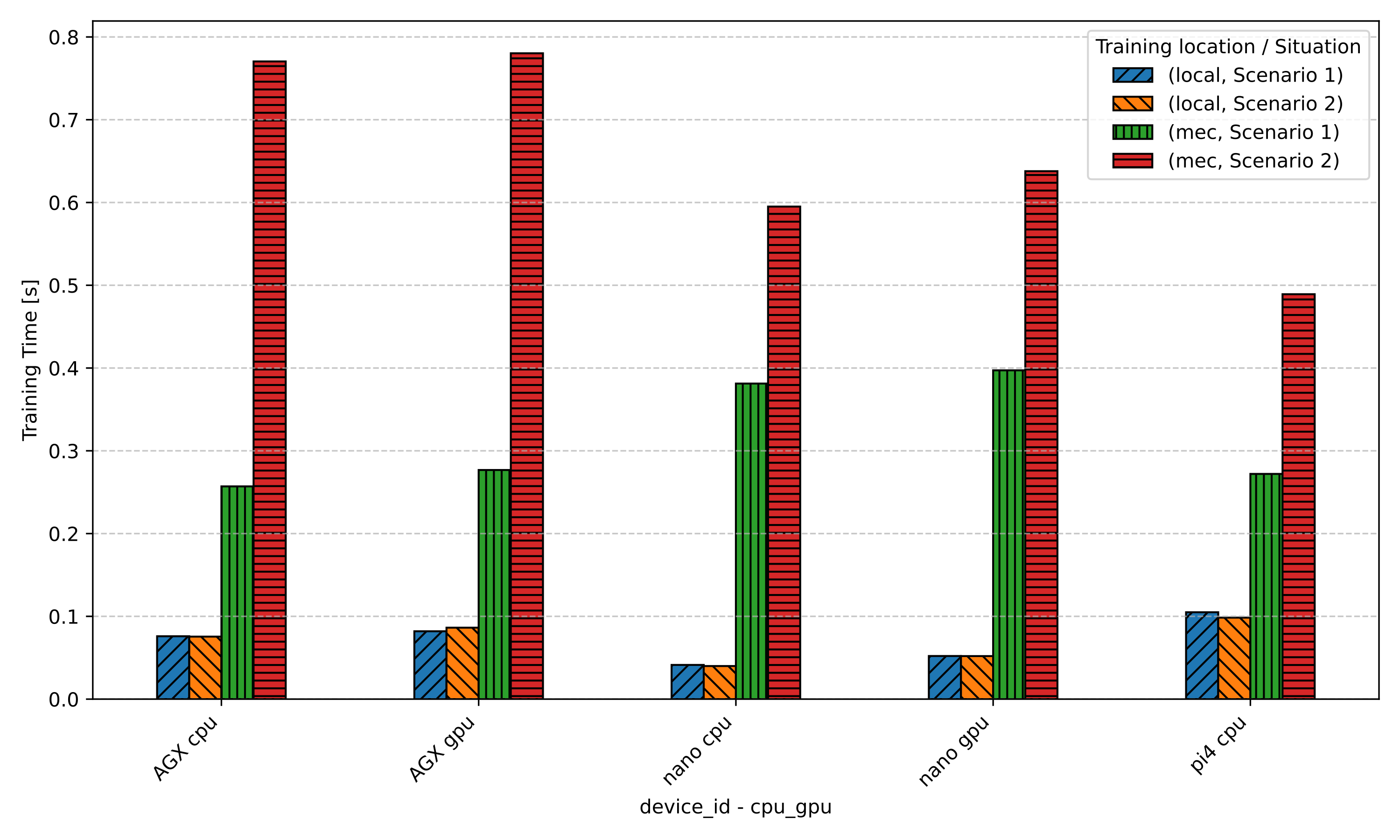}
    \caption{Comparison of training time of \ac{DQN} with and without Multi-UE}
    \label{fig:dqn_time_compare_multiue}
\end{figure}

The same analysis is performed using the \ac{DQN} algorithm in Figure~\ref{fig:dqn_time_compare_multiue}, obtaining similar results. In this case, there is no situation where using a \ac{MEC} server is preferable for training compared to training locally. 

\begin{figure}[htbp]
    \centering
    \includegraphics[width=0.8\linewidth]{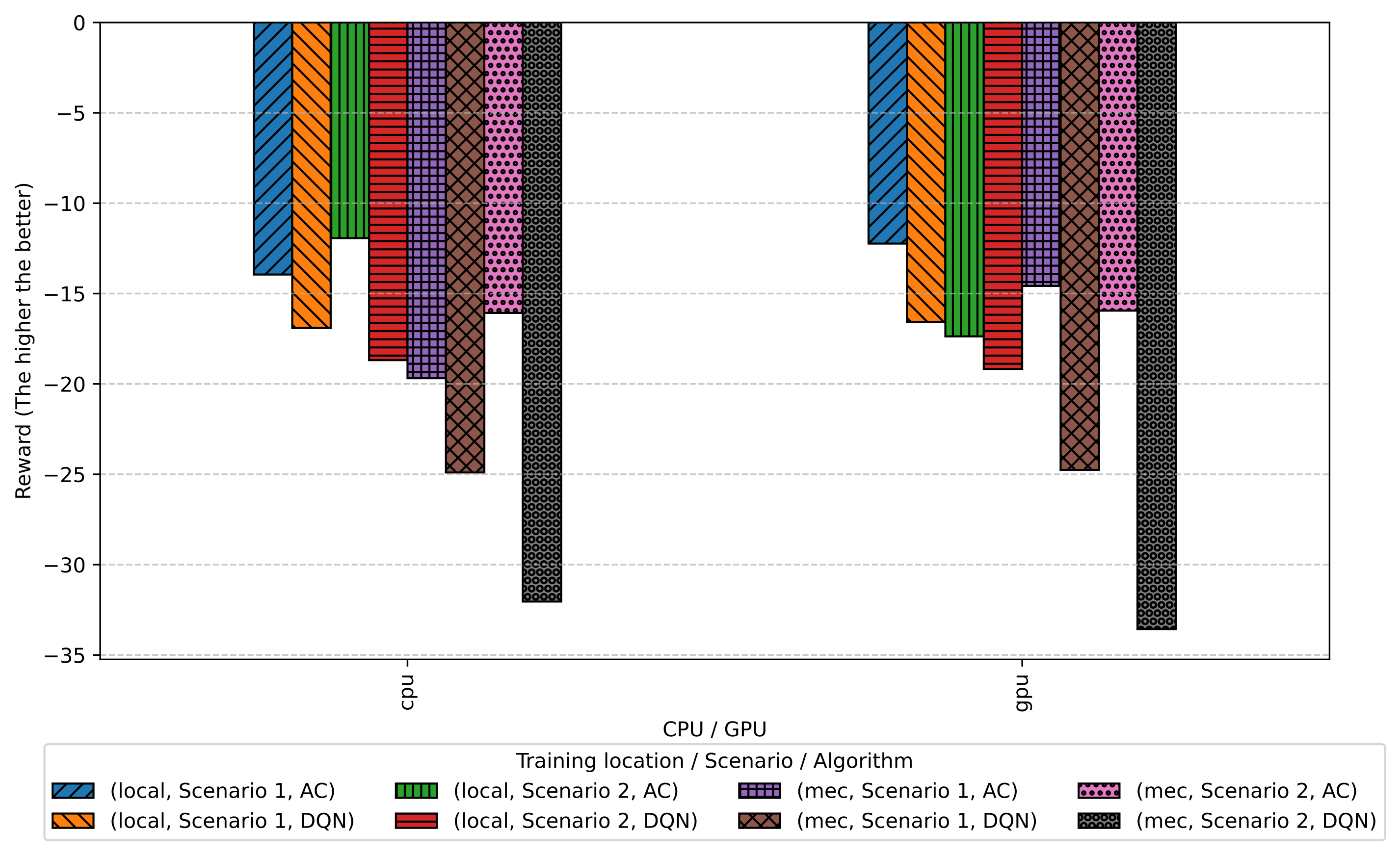}
    \caption{Rewards for AGX in each condition}
    \label{fig:overall_qoe_agx}
\end{figure}

We compare the performance on each device separately, analyzing the obtained rewards. Figure~\ref{fig:overall_qoe_agx} illustrates the performance of both algorithms using the \textit{Jetson AGX Orin} when training locally versus offloading to the \ac{MEC} server, considering both scenarios. Given the superior computational capacity of the \textit{Jetson AGX Orin}, the tendency is to favor local execution over offloading, as local training achieves higher rewards and better responsiveness. This observation reinforces the need to avoid a centralized or aggregated model, since each device exhibits distinct characteristics that influence performance. Among the algorithms, \ac{AC} consistently delivers slightly better results than \ac{DQN}, confirming its suitability for more powerful edge devices. Additionally, the results obtained using the GPU are similar to or inferior to those obtained using the CPU.

\begin{figure}[htbp]
    \centering
    \includegraphics[width=0.8\linewidth]{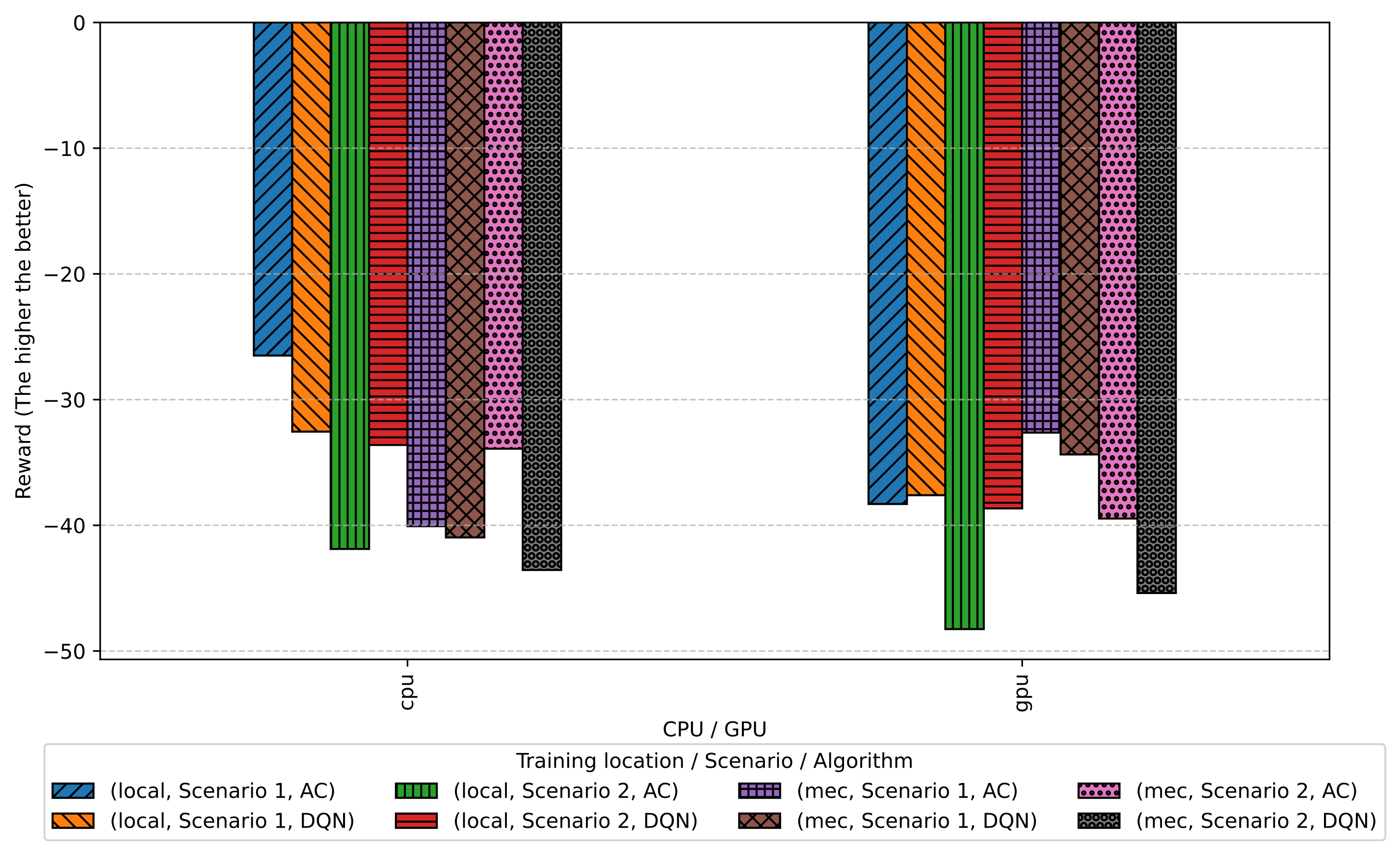}
    \caption{Rewards for reComputer J1010 in each condition}
    \label{fig:overall_qoe_nano}
\end{figure}

In the case of \textit{reComputer J1010}, which is illustrated in Figure~\ref{fig:overall_qoe_nano}, the best results are also obtained when training locally in Scenario 1 with both algorithms. However, the computation capabilities of \textit{reComputer J1010} are smaller, so the obtained rewards are worse than using the \textit{Jetson AGX Orin}. In any case, the conclusions drawn from the \textit{Jetson AGX Orin} can be extrapolated to the results when using the \textit{reComputer J1010}.

\begin{figure}[htbp]
    \centering
    \includegraphics[width=0.8\linewidth]{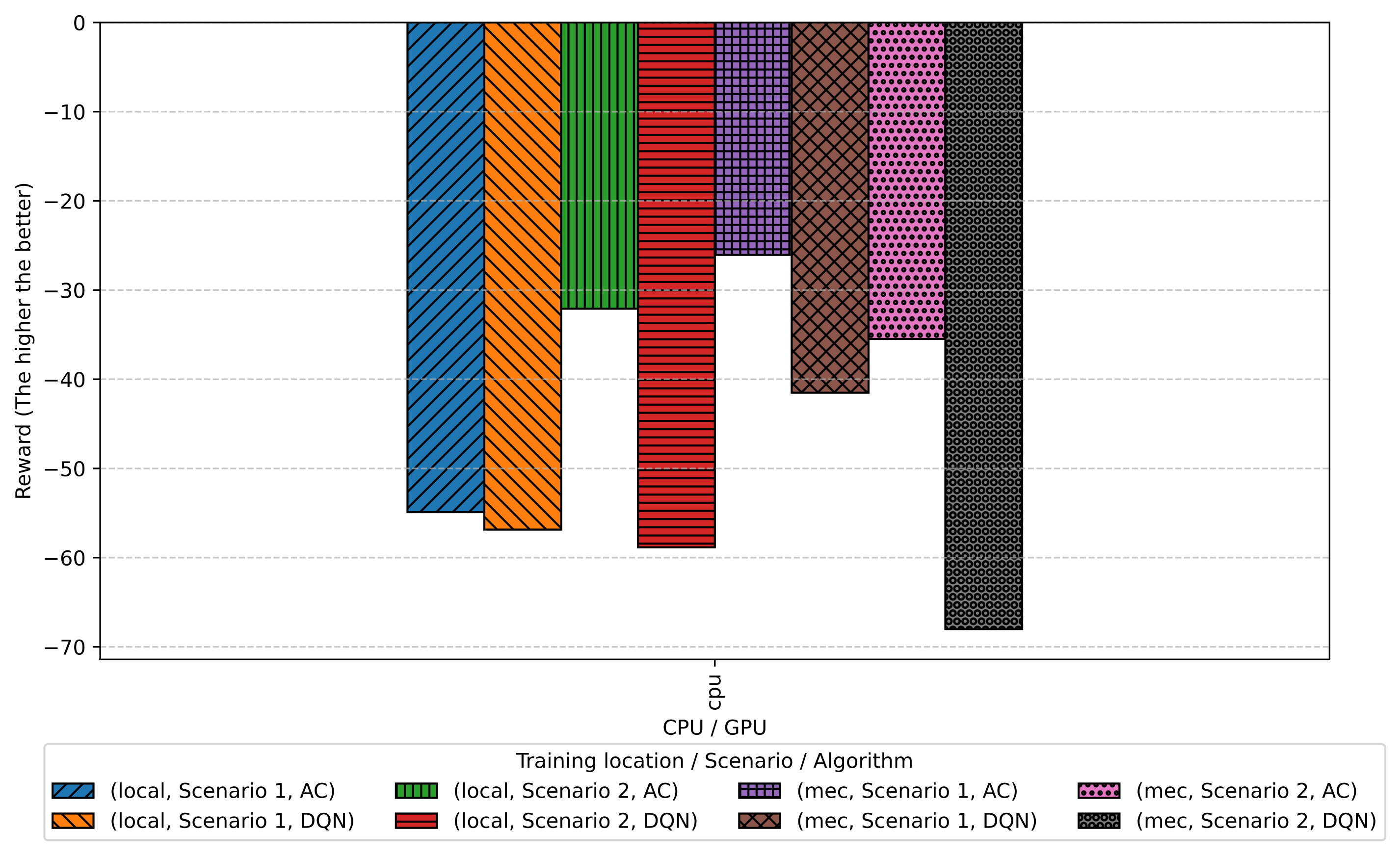}
    \caption{Rewards for Raspberry Pi in each condition}
    \label{fig:overall_qoe_raspi}
\end{figure}

Finally, as shown in Figure~\ref{fig:overall_qoe_raspi}, the \textit{Raspberry Pi} achieves the lowest reward among the evaluated devices. In Scenario 1, the \ac{AC} algorithm can't select optimal actions, likely due to the device's limited computational resources. 

The results of this Section demonstrate that the proposed \ac{DRL} agents are both effective and feasible for adaptive task offloading in edge environments. On-device training proves viable under certain constraints, particularly when leveraging energy-efficient CPU execution rather than GPU acceleration. Even on high-end platforms such as \textit{Jetson AGX Orin}, GPU usage provides negligible benefits in continual learning settings without batching, while increasing energy consumption. For lightweight online \ac{DRL} tasks, CPU-based execution or adaptive hybrid architectures appear more suitable than GPU-centric designs. Additionally, offloading training to the \ac{MEC} server offers limited advantages for these models. It introduces risks related to network latency and resource contention, depending on the environment's variability, which is one reason to apply \ac{DRL} techniques to solve the computation offloading problem.

\section{Conclusion}
\label{sec:conclusion}
This work provides several key insights into the practical deployment of \ac{DRL}-based task offloading in edge environments. In this work, we present the feasibility and benefits of running on-device \ac{DRL} to address the computation offloading problem in real-world scenarios. We implemented the whole process of different \ac{DRL} approaches on-device to solve the computation offloading problem in a fully decentralized manner, demonstrating the benefits compared to centralized solutions.

We utilize a testbed that can generate a variety of dynamic environments, changing the characteristics of the tasks, as well as the network conditions, through the number and traffic exchange of the emulated \acp{UE}. We also alter the servers' status by emulating their usage as if they were being used by different \acp{UE} or processes, which can also affect training when performed in a centralized manner, highlighting the need for on-device learning.

The results demonstrate that online, on-device \ac{DRL} training is not only feasible on modern edge platforms, but also more energy-efficient and responsive than remote training approaches. Additionally, GPU acceleration does not offer advantages for lightweight \ac{DRL} training at the edge, and in some cases, it degrades performance due to additional overhead. 

These findings have direct implications for the design of \ac{AIoT} systems. Rather than relying on remote training or GPU acceleration by default, system designers should prioritize lightweight, CPU-based \ac{DRL} agents trained locally, particularly in scenarios with variable network conditions and strict energy constraints.

Overall, the solution shows strong feasibility. The proposed \ac{DRL} agents perform well across tested scenarios, demonstrating their effectiveness for the intended application. Additionally, the computational overhead of local training is relatively low, suggesting that the approach can be deployed without incurring heavy resource demands. These findings support the practicality of the proposed method for real-world scenarios.

In future work, we will explore additional strategies to enhance on-device learning and scalability further. Promising directions include optimizing lightweight \ac{DRL} models for resource-constrained devices. The usage of \ac{FL} will also be analyzed in depth, as Federated \ac{DRL} approaches can enable collaborative learning without centralized aggregation, thereby accelerating the learning of \ac{DRL} models. Addressing concept drift in continual learning environments is also crucial for maintaining performance over time, as training can be a bottleneck in this environment. Finally, hybrid compute strategies, such as using a CPU for online adaptation and a GPU for periodic batch consolidation, can strike a balance between efficiency and accuracy in dynamic edge settings.

\begin{acronym}
    \acro{AC}{Actor-Critic}
    \acro{AI}{Artificial Intelligence}
    \acro{AIoT}{Artificial Intelligence of Things}
    \acro{BS}{Base Station}
    \acro{BW}{Bandwidth}
    \acro{CC}{Cloud Computing}
    \acro{CS}{Cloud Server}
    \acro{DGEMM}{Double-precision GEneral Matrix Multiplication}
    \acro{DNN}{Deep Neural Network}
    \acro{DQN}{Deep Q-Network}
    \acro{DRL}{Deep Reinforcement Learning}
    \acro{FL}{Federated Learning}
    \acro{IoT}{Internet-of-Things}
    \acro{MDP}{Markov Decision Process}
    \acro{MEC}{Multi-access Edge Computing}
    \acro{ML}{Machine Learning}
    \acro{MQTT}{Message Queuing Telemetry Transport}
    \acro{MSE}{Mean Square error}
    \acro{NN}{Neural Network}
    \acro{NR}{New Radio}
    \acro{ONNX}{Open Neural Network eXchange}
    \acro{ReLU}{Rectified Linear Unit}
    \acro{RL}{Reinforcement Learning}
    \acro{SA}{Stand-Alone}
    \acro{SGEMM}{Single-precision GEneral Matrix Multiplication}
    \acro{SL}{Supervised Learning}
    \acro{TanH}{Hyperbolic Tangent}
    \acro{TD}{Temporal Difference}
    \acro{UE}{User Equipment}
    \acro{WCET}{Worst-Case Execution Time}
\end{acronym}

\bibliography{export.bib}{}
\bibliographystyle{IEEEtran}

\newpage

\vfill

\end{document}